\documentclass[]{jfm}
\usepackage{graphicx}
\usepackage{color}
\usepackage{amssymb,amsfonts,amsmath}
\usepackage{gensymb}
\usepackage{overpic}
\usepackage{wasysym}

\newcommand{\ve}[1]{\ensuremath{\boldsymbol{#1}}}
\newcommand{\sve}[1]{\ensuremath{\boldsymbol{\footnotesize#1}}}
\newcommand{\ma}[1]{\ensuremath{\mathsfbi{#1}}}

\newcommand{\tr}{\ensuremath{\mbox{Tr}}}

\newcommand\nn{\nonumber}

\DeclareMathOperator{\ku}{Ku}
\DeclareMathOperator{\st}{St}
\DeclareMathOperator{\abar}{{\overline{a}}}

\newcommand{\eqnlab}[1]{\label{eq:#1}}
\newcommand{\figlab}[1]{\label{fig:#1}}

\newcommand{\eqnref}[1]{\eqref{eq:#1}}
\newcommand{\figref}[1]{\ref{fig:#1}}

\newcommand{\Eqnref}[1]{Eq.~\eqref{eq:#1}}
\newcommand{\Figref}[1]{Fig.~\ref{fig:#1}}
\newcommand{\Tabref}[1]{Table~\ref{tab:#1}}

\newcommand{\Secref}[1]{Sec.~\ref{sec:#1}}

\newcommand{\bx}{{\ve x}}
\newcommand{\bk}{{\ve k}}
\newcommand{\bq}{{\ve q}}

\newcommand{\ccdot}{\boldsymbol{\cdot}}

\newcommand{\T}{^{\rm T}}

\begin{document}

\title{Helicoidal particles in turbulent flows with multi-scale helical injection\footnote{Postprint version of the article published in Journal of Fluid Mechanics {\bf 869} 646--673 (2019) DOI:10.1017/jfm.2019.237
}
}
\author{L. Biferale\aff{1},
  K. Gustavsson\aff{2}
  \corresp{\email{kristian.gustafsson@physics.gu.se}}
   \and R. Scatamacchia\aff{1}}

\affiliation{\aff{1}Department of Physics and INFN, University of Rome Tor Vergata,
Rome, 00133, Italy
\aff{2}Department of Physics, University of Gothenburg,
Gothenburg, 41296, Sweden

}

\maketitle

\begin{abstract}
We present numerical and theoretical results concerning the properties of turbulent flows with strong multi-scale helical injection. We perform direct numerical simulations of the Navier-Stokes equations   under a random helical stirring with power-law spectrum  and with different intensities of energy and helicity injections. We show that there exists three different regimes where the forward energy and helicity inertial transfers are: (i) both leading with respect to the external injections, (ii) energy transfer is leading and helicity transfer  is sub-leading and (iii) both are sub-leading and helicity is maximal at all scales. As a result, the cases (ii-iii) give flows with Kolmogorov-like inertial energy cascade and tuneable helicity transfers/contents. We further explore regime (iii) by studying its effect on the kinetics of point-like isotropic helicoids, particles whose dynamics is isotropic but breaks parity invariance.
We investigate small-scale fractal clustering and preferential sampling of intense helical flow structures. Depending on their structural parameters, the isotropic helicoids either preferentially sample co-chiral or anti-chiral flow structures. We explain these findings in limiting cases in terms of what is known for spherical particles of different densities and degrees of inertia. Furthermore, we present theoretical and numerical results for a stochastic model where dynamical properties can be calculated using analytical perturbation theory. Our study shows that a suitable tuning of the stirring mechanism can strongly modify the small-scale turbulent helical properties and demonstrates that isotropic helicoids are the simplest particles able to preferentially sense helical properties in turbulence.

\end{abstract}

\section{Introduction}
\noindent
Helicity is an invariant of the Navier-Stokes equations (NSE) in three spatial dimensions when neglecting the effects of viscous dissipation and external forcing  \citep{Frisch,moffatt1992helicity,chen2003joint,alexandros2018}. It is  connected to the topological structure of vortex lines,  characterized in terms of  twist, writhe and linking numbers \citep{scheeler2014helicity,kedia2016weaving,laing2015conservation,kerr2015simulated}. Helicity can be introduced in a flow by a stirring mechanism that breaks mirror symmetry and its effects on the turbulent energy cascade in three spatial dimensions have been widely studied since the pioneering work of \citet{brissaud1973} (see  \citet{borue1997spectra, pelz1985velocity, kerr1987histograms, kholmyansky1991some, kit1987experimental} for other  contributions). In geophysical flows, helicity plays an important role in the  atmospheric Ekman layer, where there exist arguments supporting a turbulent helicity cascade in the logarithmic range of the boundary layer \citep{deusebio2014helicity,koprov2005experimental,kurgansky2017helicity}. Recent experimental advancements allowed the production of vortex bundles with a different prescribed topology  \citep{kleckner2013creation} and the combination of shear and helicity has been studied experimentally and numerically \citep{herbert2012dual,qu2018cascades}.
Concerning the dual energy-helicity cascade, it is widely believed that for the case of NSE in three spatial dimensions forced on a limited range of scales, both energy and helicity cascade forward \citep{chen2003joint,chen2003intermittency,sahoo2015}. This is a dual co-directional cascade according to the classification given in \citet{alexandros2018}. The mirror-symmetry breaking induced by the helical stirring mechanism tends to become weaker and weaker by going to smaller and smaller spatial scales: the energy transfer is {the leading mechanism} and small-scale turbulence recovers a neutral statistics with zero helicity on average. On the contrary, if only one homochiral sector is dynamically active, one can prove that NSE admit a dual counter-directional  cascade (with energy flowing backward and helicity forward). For this case the flow has global solutions \citep{waleffe1992nature,biferale2012,biferale2013global} and small-scale turbulence is strongly (maximally) helical. In addition there are analytical and numerical hints \citep{linkJFM} that helicity induces a non-trivial decrease in the drag coefficient of turbulent flows.

\noindent
In this paper we further investigate the statistical properties of the dual energy-helicity transfers by adopting a power-law multi-scale  stirring mechanism, which allows us to explore three different regimes concerning the relative intensity of energy and helicity injections. In particular, we show that there exists a suitable range of forcing spectral exponents, where the energy transfer is not affected by the stirring term while helicity can be controlled, leading to a turbulent realization with tuneable small-scale helicity content. Furthermore, in a regime where both small-scale energy and helicity contents are controlled by the forcing, leading to  maximal-helicity flow configurations,
we study the preferential concentration of isotropic helicoids \citep{Kel71,Gus16a}, i.e. point-like particles whose dynamics is isotropic but breaks mirror symmetry. By using both direct numerical simulations (DNS) and a stochastic model for the Eulerian advecting velocity field \citep{Gus16b}, we show that isotropic helicoids possess highly non-trivial preferential sampling of the underlying helical flow properties depending on the particle parameters.
The paper is organized as follows.
In \Secref{Flow} we describe the Eulerian part, discussing the different regimes for different helical injection power spectra and we present numerical simulations of the different regimes.
In \Secref{Helicoids} we introduce the isotropic helicoids and their dynamical equations. We discuss the existence of two new scales of the Stokes number, $\st_{\pm}$, which depend on the coupling between translational and rotational degrees of freedom. Furthermore,  we present results on the preferential sampling of the flow helicity for different particle parameters, including two asymptotic limits where the Stokes number $\st$ is either much smaller than $\st_+$ or much larger than $\st_-$.
We conclude the paper in \Secref{Conclusion}.

\section{{Helical turbulent flows: Eulerian properties}}
\label{sec:Flow}

\noindent
\subsection{Theoretical background}
\noindent
We start by considering the forced NSE for the fluid velocity $\ve u$ and the pressure $p$ in three spatial dimensions:
\begin{equation}
\label{eq:nse}
\partial_t {\ve u} + {\ve u} \ccdot {\ve \nabla} {\ve u} = - {\ve \nabla} p + \nu \Delta {\ve u} + {\ve f}\;,\hspace{1cm} \ve\nabla\ccdot\ve u=0\,,
\end{equation}
where $\nu$ is the kinematic viscosity and ${\ve f}$ is
a parity-breaking external forcing with energy injection {rate} $\epsilon = \langle {\ve u}\ccdot{\ve f} \rangle$
and helicity injection {rate} $h = \langle {\ve u}\ccdot( {\ve \nabla} \times {\ve f}) + 2 {\ve \Omega}\ccdot {\ve f} \rangle$, where  $2 {\ve \Omega} = {\ve \nabla} \times {\ve u}$  denotes the flow vorticity.
It is useful to adopt an exact decomposition of the velocity field in positive and negative Fourier helical waves \citep{constantin1988beltrami,waleffe1992nature}:
\begin{equation}
\label{eq:hel_waves}
{\ve u}({\ve x}, t) = \sum_k [ u_\bk^+(t){\ve h}_\bk^+ + u_\bk^-(t){\ve h}_\bk^- ]e^{-i \bk \cdot \bx}\,,
\end{equation}
\noindent
where ${\ve h}_k^{\pm}$ are the eigenvectors of the curl operator. In terms
of such decomposition the total energy, $E = \int{\rm d}^3x\,\ve u^2$, and the total helicity, {$H=2\int {\rm d}^3x\,\ve u\ccdot\ve\Omega$}, take the forms:
\begin{equation}
\label{eq:energy}
E = \sum_\bk |u_\bk^+|^2 + |u_\bk^-|^2\;, \qquad
H = \sum_\bk k( |u_\bk^+|^2 - |u_\bk^-|^2)\,.
\end{equation}
\noindent
We can further consider the energy content of positive and negative helical modes,
$E^{\pm}(k) = \sum_{ |{\ve k}| = k} |u_\bk^{\pm}|^2$, where $\Delta k = 2\pi/L$, such that the energy
and helicity spectra become
\begin{equation}
\label{eq:en_scale}
E(k) = E^+(k) + E^-(k)\;, \qquad
H(k) = k[E^+(k) - E^-(k)]\,.
\end{equation}
Supposing that there exists a dual co-directional forward cascade of energy and helicity and that the typical time at scale $ r \sim k^{-1}$ is dominated by the energy eddy turnover time
$\tau_E(r) \sim \epsilon^{-1/3} r^{2/3}$,  we have for the semi-sum and semi-difference of the
spectral components \citep{chen2003joint}:
\begin{equation}
\label{eq:semi_sum}
 E^+(k) + E^-(k) \sim C_E \epsilon^{2/3} k^{-5/3}\;, \qquad
E^+(k) - E^-(k) \sim C_H h \epsilon^{-1/3} k^{-8/3}\,,
\end{equation}
where $C_E$ and $C_H$ are {two constants of dimension inverse length}. Hence the two energy components can be written as:
\begin{equation}
\label{eq:en_comp}
E^{\pm}(k) \sim C_E \epsilon^{2/3} k^{-5/3} \pm C_H h \epsilon^{-1/3} k^{-8/3}\,.
\end{equation}
\noindent
It is known that for large-scale energy and helicity injection the Kolmogorov-like scaling (\ref{eq:semi_sum}) is observed, implying a recovery of mirror symmetry at small scales, see for example \cite{sahoo2015,Val18} for recent studies about this issue with and without rotation. In order to have strong multi-scale helicity, it is necessary to resort to a power-law injection \citep{Forst97,Seoud07}.
\noindent
\subsection{Multi-scale energy and helicity injections}
\noindent
Let us suppose a Gaussian white-in-time helical forcing, $${\ve f}({\ve x},t) = \sum_\bk  f_\bk^+(t){\ve h}_\bk^+ e^{-i \bk \cdot \bx},$$ whose two-point correlation is isotropic, and with a power-law spectrum \citep{sain1998turbulence,biferale2004effects,kessar2015non}:
\begin{equation}
\langle f^+_\bk(t)  f^+_{\bk'}(t') \rangle = D_0 k^{1-d-y} \delta(t-t') \delta_{\bk,\bk'}\,,
\eqnlab{forcing_helical}
\end{equation}
where $d$ is the space dimension and $D_0$ defines the typical forcing intensity at the smallest wavenumber that we will always assume to be $k_0=2\pi/L=1$.
For the sake of numerical implementation we cut off the power-law at a maximum  wavenumber of the order of the Kolmogorov scale, $k_{max} \sim k_\eta$.
Using this forcing, the energy and helicity injection rates up to the scale $k < k_{max}$  can be estimated as:
\begin{equation}
\label{eq:en_fft}
\epsilon(k) \sim  \sum_{|\bq|<k} |\bq|^{1-d-y}\;, \qquad h(k) \sim   \sum_{ |\bq|<k}|\bq|^{2-d-y}\,.
\end{equation}
By considering spherical symmetry, the sums in (\ref{eq:en_fft}) can be easily estimated and we distinguish three different regimes depending on the forcing spectrum: (I) when $ y >2$ both energy and helicity injections are dominated by the infrared range, $\epsilon(k) \to const. $ and $h(k) \to const. $  when $k \to \infty$, and the system behaves as if it is forced at large scales only. In this case, we obtain a dual energy-helicity cascade because both quantities are transferred by the nonlinear inertial terms of the NSE (\ref{eq:nse}); (II) when $1<y<2$ the energy injection sum is still dominated by the infrared range, while the helicity injection depends on the ultra-violet limit, $h(k) \sim k^{2-y}$. In this regime we obtain an energy cascade and helicity multi-scale injection; (III) when $ y <1$ both energy and helicity transfer are dominated by the multi-scale injection, $\epsilon(k) \sim k^{1-y}$ and $h(k) \sim k^{2-y}$. The three regimes are summarized in \Tabref{regimes}.
\begin{table}
\begin{center}
\begin{tabular}{ccc}
(I) & (II) & (III) \\[3pt]
$y > 2$ & $1 < y < 2$ & $y < 1$ \\
$\epsilon(k) = const.$ & $\epsilon(k) = const.$ & $\epsilon(k) \sim k^{1-y}$ \\
$h(k) = const.$ & $h(k) \sim k^{2-y}$ & $h(k) \sim k^{2-y}$
\end{tabular}
\caption{\label{tab:regimes} Energy and helicity injection regimes depending on the forcing spectrum in \Eqnref{forcing_helical}.}
\end{center}
\end{table}

\noindent
As a result, the spectral properties (\ref{eq:en_comp}) are valid only for regime (I), and we can summarize the scaling for all three different  regimes as follows:
\begin{equation}
\begin{cases}
\label{eq:3spectra}
E^{\pm}(k) \sim C^I_E \epsilon^{2/3} k^{-5/3} \pm C^I_{H} h \epsilon^{-1/3} k^{-8/3}\;,\qquad y > 2\\
E^{\pm}(k) \sim C^{II}_E \epsilon^{2/3} k^{-5/3} \pm C^{II}_{H} k^{2-y}\epsilon^{-1/3} k^{-8/3}\;,\qquad 1 <y <2\\
E^{\pm}(k) \sim C^{III}_E k^{2(1-y)/3} k^{-5/3} \pm C^{III}_{H} k^{2-y}k^{-(1-y)/3} k^{-8/3}\;,\qquad y <1\,,
\end{cases}
\end{equation}
where the prefactors depend on the forcing intensity \eqnref{forcing_helical}.
From the expressions \eqnref{3spectra} we can evaluate the mirror-symmetry recovery ratio, $R(k) = |E^+(k)-E^-(k)|/(E^+(k)+E^-(k))$ in the three regimes as:
\begin{equation}
\begin{cases}
\label{eq:recovery}
R^I(k) \sim k^{-1}\;,\qquad y > 2\\
R^{II}(k) \sim  k^{1-y}\;, \qquad 1 <y <2\\
R^{III}(k) \sim const.\;, \qquad y <1\,,
\end{cases}
\end{equation}
from which it follows that regime (III) is a flow with a maximal helical content at all scales where the injection is acting. Before concluding this section it is important to stress again that the prediction leading to regime III is obtained under the assumption that the typical time scale guiding the transfer is the scale-dependent  generalization of the eddy turnover time: $\tau_E(k) \propto k^{-2/3}\epsilon(k)^{-1/3}$, which is not necessarily the  only possibility. In order to have a quantitative assessment of the scaling properties at high Reynolds numbers one could resort to Fourier closures based on eddy-damped quasi-normal Markovian (EDQNM) approximation as in \citet{briard2017}.
In the following, we  resort to direct numerical simulations and we present a first numerical investigation of the flow properties under multi-scale helical injection without any approximation.
\subsection{Numerical simulation}
\label{sec:numerical}
\begin{figure}
\begin{center}
\includegraphics[width=13.8cm,draft=false]{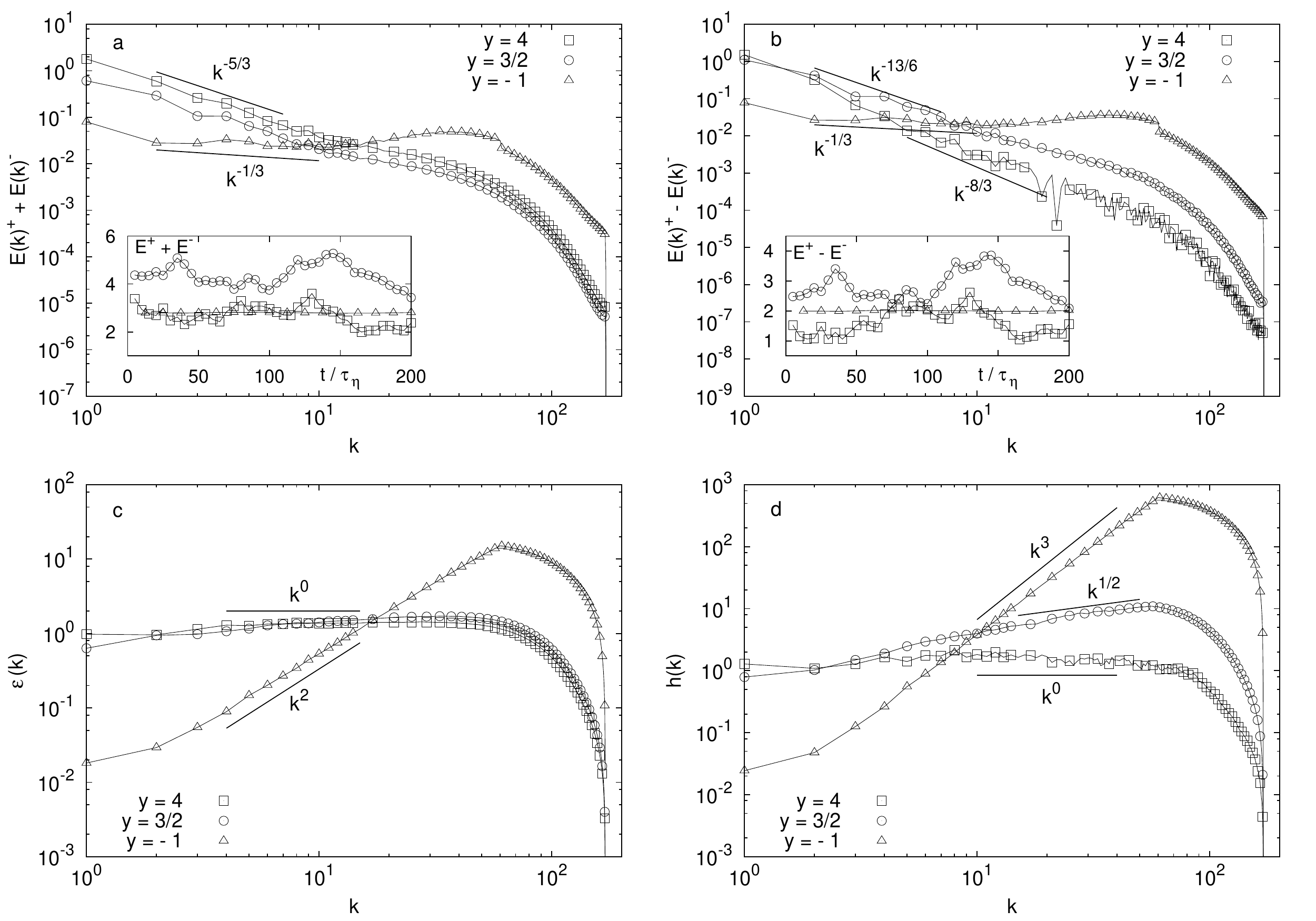}
 \caption{Time average of the energy and helicity spectra ({\bf a},{\bf b}) and fluxes ({\bf c},{\bf d}) for the three regimes $y=4$ (I), $y=3/2$ (II), and $y=-1$ (III). The small discontinuity at the high wave numbers is due to the end of the range where the forcing is applied. Inset: Time evolution of the total energy ({\bf a}) and total helicity ({\bf b}) in the stationary regime where all averages are performed.
Parameters are given in \Tabref{param}.
In panel {\bf a}, the curve for $y=3/2$ ($\Circle$) has been shifted with respect to the curve for $y=4$ ($\square$) for the sake of presentation. We also superpose the scalings predicted by the relations in Eqs. (\ref{eq:3spectra}) and (\ref{eq:en_fft}).}
 \label{fig:1}
\end{center}
\end{figure}

\begin{figure}
\begin{center}
\includegraphics[width=13.8cm,draft=false]{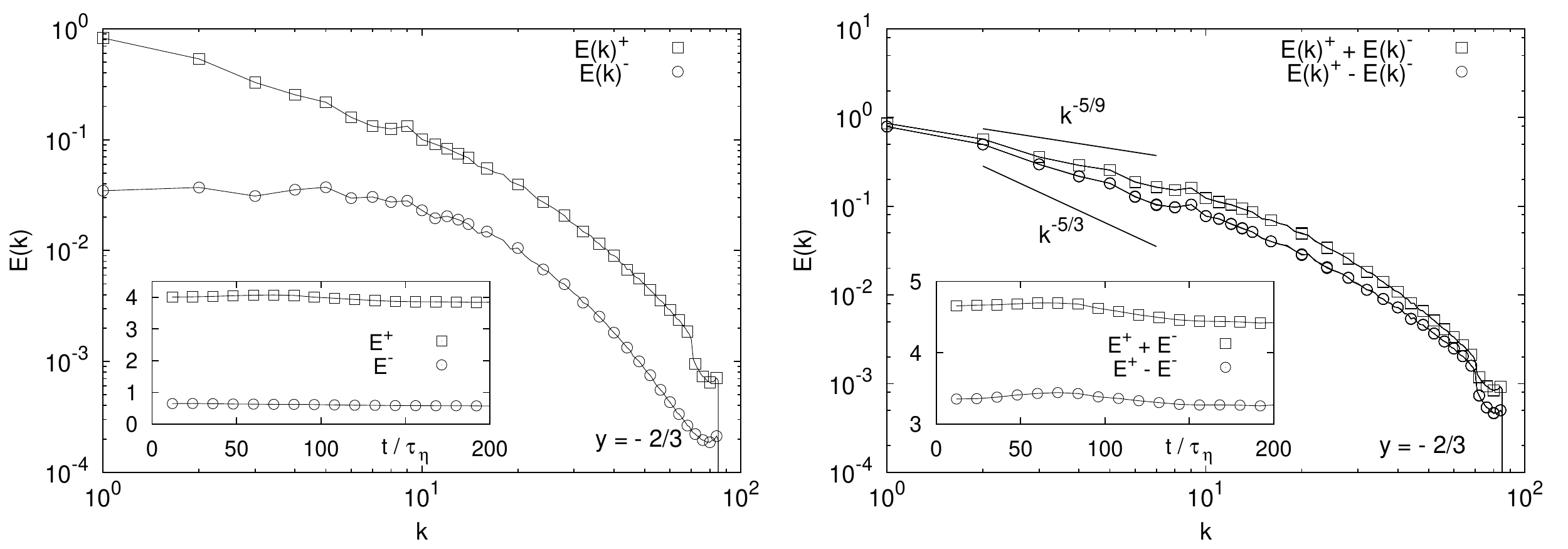}
 \caption{Left: Time average of positive and negative helical spectral components for the direct numerical simulations with parameters of the fourth parameter set in \Tabref{param}. Inset: Time evolution of the helical spectral components in the stationary regime.
Right: Total energy $E^+(k)+E^-(k)$ and rescaled total helicity $E^+(k)-E^-(k) = H(k)/k$. The scaling $-5/9$ predicted by relation (\ref{eq:3spectra}) and the Kolmogorov $-5/3$ power laws are also shown for comparison. Inset: Time evolution of total energy and rescaled total helicity.}
 \label{fig:2}
\end{center}
\end{figure}

\noindent
In this section we show the results of a series of DNS with resolution of $512^3$ grid points to
explore properties of the energy and helicity of the three fluid regimes identified in the previous section. We implement a hyper-viscosity
method to extend the inertial range \citep{borue1995self}. In particular we set $\nu_\alpha \Delta^\alpha u$ as the viscous term,
with $\alpha = 2$.
The external forcing ${\ve f}$ in  \eqnref{nse} has been implemented as a Langevin process with correlation time proportional to a fraction of the Kolmogorov time.
As detailed in the previous section, to obtain a fully helical flow, we project the forcing only on velocity modes with positive helicity with energy injection at all wavenumbers {up to} dissipative scales $k \in [1:70]$. Three representative values for the three regimes have been selected: $y = 4, 3/2, -1$.
Details about the $512^3$ DNS set-ups are summarized in \Tabref{param}. In \Figref{1}, we present four panels with the results for ({\bf a}) the energy spectrum, ({\bf b}) the helicity spectrum, ({\bf c}) the energy flux and ({\bf d}) the helicity flux as functions of $k$ and for the three representatives values of $y$. In the insets of panels ({\bf a}, {\bf b}) the total energy and helicity as functions of time in the stationary regime are shown. The predictions for spectra (\ref{eq:3spectra}) and  energy fluxes (\ref{eq:en_fft}) are verified with good accuracy, except for ultraviolet effects induced by the cutoff wavenumber where we stop to act with the external forcing to avoid stability issues in the code.
Note that the power-law forcing smooths down the presence of the high-wavenumber bottleneck expected in the spectrum when using hyper-viscosity \citep{bottle}.
Overall, we conclude that by changing the spectral properties of the helically forced NSE we can achieve a flow evolution with tuneable energy/helicity ratios as theorized by (\ref{eq:recovery}). In particular, in \Figref{2}({\bf a}) we show both the positive and negative helical spectral components $E^\pm(k)$ for {$y=-2/3$} (case III). The major contribution to the energy spectrum is given by the velocity modes with positive helicity $E^+(k)$ for all wavenumbers.
As a result the Navier-Stokes flow develops a dominant positive helical dynamics at all scales.
The \Figref{2}({\bf b}) shows that $|E^+(k)-E^-(k)| \sim (E^+(k)+E^-(k))$ which implies that mirror symmetry is broken at all scales.
We remark that in this regime, the scaling behaviour of $|E^+-E^-|$ and $E^++E^-$ are less steep than the Kolmogorov prediction that is in both cases dominated by the external injection as predicted by (\ref{eq:3spectra}).

\begin{table}
  \begin{center}
  \begin{tabular}{lcccccccccc}
Regime & $N^3$ & $\eta$ & $\Delta x$ & $\Delta t$ & $\nu$ & $\tau_\eta$ & $\tau_S$ & $y$ & $\alpha$ & $N_h$\\[3pt]
(I) &  $512^3$ & 0.008 & 0.012 & 0.0003 & 1.9 $\times 10^{-7}$ & $0.03$ & $1$ & 4 & 2 & -\\
(II) &  $512^3$ & 0.008 & 0.012 & 0.0003 & 1.9 $\times 10^{-7}$ & $0.03$ & $1$  & 3/2 & 2 & -\\
(III) &  $512^3$ & 0.008 & 0.012 & 0.0003 & 1.9 $\times 10^{-7}$ & $0.012$ & $1$  & $-1$ & 2 & -\\
(III) &   $256^3$ & 0.016 & 0.024 & 0.0006 & 0.0052 & $0.025$ & $500$ & ${-2/3}$ & 1 & $2.4 \times 10^{6}$
  \end{tabular}
  \caption{Parameters of the numerical simulations: grid resolution $N^3$, Kolmogorov
    length scale $\eta$ in simulation units (SU), grid spacing $\Delta x = 2 \pi/N$ (SU), time step $\Delta t$ (SU), kinematic viscosity $\nu$ (SU), Kolmogorov eddy turn-over time $\tau_\eta = (\nu/ \epsilon)^{1/2}$ with $\epsilon$ the energy dissipation rate (SU), forcing correlation time $\tau_S$ (in units of $\Delta t$), forcing power law exponent $y$, hyper-viscosity parameter $\alpha$, number of  helicoids per each family $N_h$.}
\label{tab:param}
  \end{center}
\end{table}

\subsection{Stochastic helical flows}
\label{sec:stochastic}
\begin{figure}
\begin{center}
\begin{overpic}[width=6cm,draft=false]{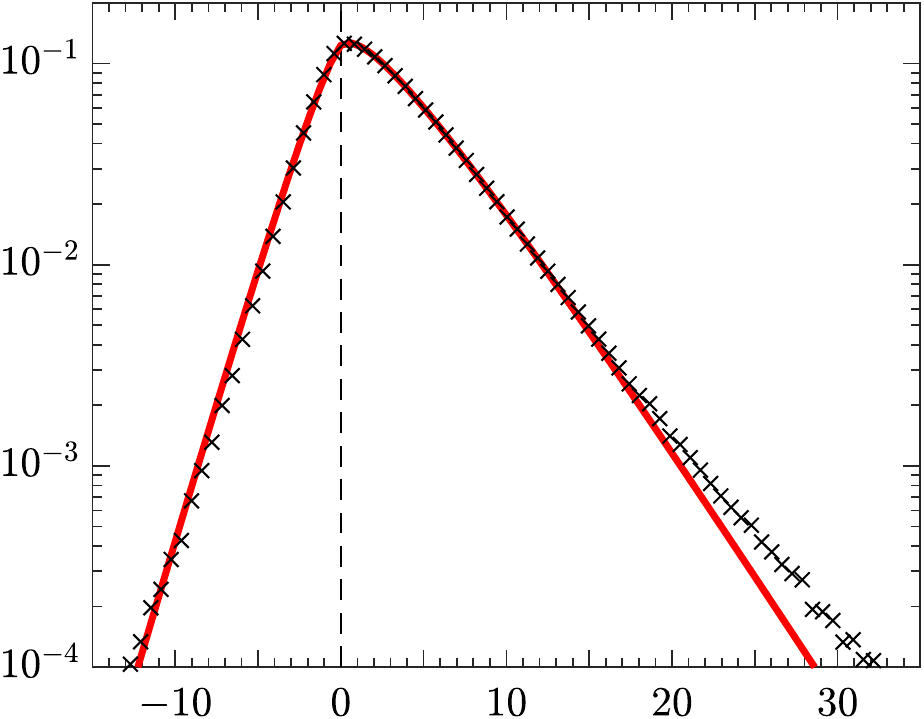}
\put(54,-5){$H$}
\put(-6,35){\rotatebox{90}{$P_0(H)$}}
\end{overpic}
\vspace{0.2cm}
\caption{Distribution of helicity of the flow $P_0(H)$ for DNS with parameters given by the fourth case in \Tabref{param} (black crosses) and for the stochastic model \Eqnref{PH0} with $H_0=0.85$ (red line).
The helicity is made dimensionless using the Kolmogorov scales $\eta$ and $\tau_\eta$.
}
\label{fig:FlowDistH}
\end{center}
\end{figure}

\noindent
The fully helical flow described by the regime (III) can be considered a sort of multi-scale flow dominated by the external forcing, where the Navier-Stokes nonlinear evolution is sub-leading with respect to the forcing effects at all scales. In order to have an analytical control and variability of the governing flow, we study also surrogate dynamics given by simpler stochastic evolution without any underlying structure coming from NSE. This approximation is also necessary to perform analytical estimates for the dynamics of particles in the flow as discussed later. To follow this idea, we consider a random incompressible, homogeneous and isotropic single-scale velocity field, $\ve u=\ve\nabla\times\ve A$.
Here the components of the vector potential $\ve A(\ve x,t)$ are independent Gaussian random functions with zero mean,
a spatial correlation function decaying on a scale of order $\eta_0$ and an exponential time-correlation function with decay rate, $\tau_0$ (see Appendix \ref{sec:StochasticModel_SI} for more details).
The velocity field is normalized such that $\langle\ve u^2\rangle=u_0^2$.
The flow is characterized by a dimensionless Kubo number
\begin{align}
\ku = u_0\tau_0/\eta_0\,,
\eqnlab{Kubo}
\end{align}
the ratio between the Eulerian flow decorrelation time $\tau_0$ and the advecting time, $\eta_0/u_0$.
The Kubo number can be seen as a dimensionless correlation time of the flow.
If $\ku$ tends to zero a white-noise flow is approached and if $\ku$ is large a persistent flow is obtained.
The latter case is important because the particle dynamics often agrees qualitatively or even quantitatively with the dynamics in a real turbulent flow~\citep{Gus15,Gus16b,Gus17}.
The former case is important because it allows for an analytical perturbative analysis in the Kubo number~\citep{Gus11,Gus16b}, and to understand the particle dynamics quantitatively at small $\ku$ and qualitatively at large  $\ku$ or in DNS.
In order to control the probability distribution function of the parity-breaking structures in
the flow, we adopt the exact helical decomposition of each Fourier mode given by (\ref{eq:hel_waves}).
Weighting the positive modes $\ve h^+_{\sve k}$ with a factor $\mu$ leads to flows where positive ($\mu>1$, $\langle H\rangle_{\rm flow}>0$) or negative ($\mu<1$, $\langle H\rangle_{\rm flow}<0$) helical structures are dominant.
The resulting flow has the following exponential-like distribution of helicity (see Appendix \ref{sec:StochasticModel_SI} for details):
\begin{align}
P_0(H) & = \frac{9}{\pi}\frac{\eta_0^2}{u_0^4}\frac{|H|\exp\left[\frac{3H_0 }{5-H_0^2}\frac{H\eta_0}{u_0^2}\right]K_1\left[\frac{3\sqrt{5}}{5-H_0^2}\frac{|H|\eta_0}{u_0^2}\right] }{\sqrt{5\left[5-H_0^2\right]}}\,,
\eqnlab{PH0}
\end{align}
where $K_\nu(x)$ is the modified Bessel function of the second kind and $H_0$ is the average dimensionless helicity
\begin{align}
H_0\equiv\frac{\eta_0}{u_0^2}\langle H\rangle_{\rm flow}=\frac{8}{3}\sqrt{\frac{2}{\pi}}\frac{\mu^2-1}{\mu^2+1}\,.
\end{align}
\Figref{FlowDistH} shows a comparison to $256^3$ DNS (fourth case in \Tabref{param}) using $H_0=0.85$ ($\mu\approx 1.5$) to make the shape of the distribution \eqnref{PH0} similar to that of the DNS described above.
In order to compare to DNS, it is necessary to take into account that the smooth length scale of the dissipation range in DNS is larger than the Kolmogorov length by a factor proportional to $\sqrt{{\rm Re}_\lambda}$ for not too large ${\rm Re}_\lambda$~\citep{Cal09}.
In our DNS we have ${\rm Re}_\lambda\sim 100$ and we therefore use $\eta_0\sim 10\eta_{\rm K}$ for the comparison.
We observe that the distributions in \Figref{FlowDistH} agree well for small values of $H$, but slightly disagree in the right tail. This is not surprising, we cannot expect to reproduce the exact shape of the helicity distribution in NSE with a single-scale stochastic flow.

\section{Helical turbulent  flows: suspensions of helicoidal particles}
\label{sec:Helicoids}
\noindent
The helical {flows} described in \Secref{Flow} {break} parity invariance (chiral symmetry): in configurations dominated by positive helicity, as the flow in \Figref{FlowDistH}, structures where the flow velocity and vorticity align are dominant.
Heavy, inertial spherical particles are not able to distinguish the chirality of the underlying flow, they centrifuge out of vortex structures independent of their sign of helicity.
We therefore study the dynamics of so-called isotropic helicoids~\citep{Kel71,Hap12}.
These are the simplest idealized generalization of spherical particles, their dynamics breaks parity, but remains isotropic.
\begin{figure}
\begin{center}
\begin{overpic}[width=13.5cm,draft=false]{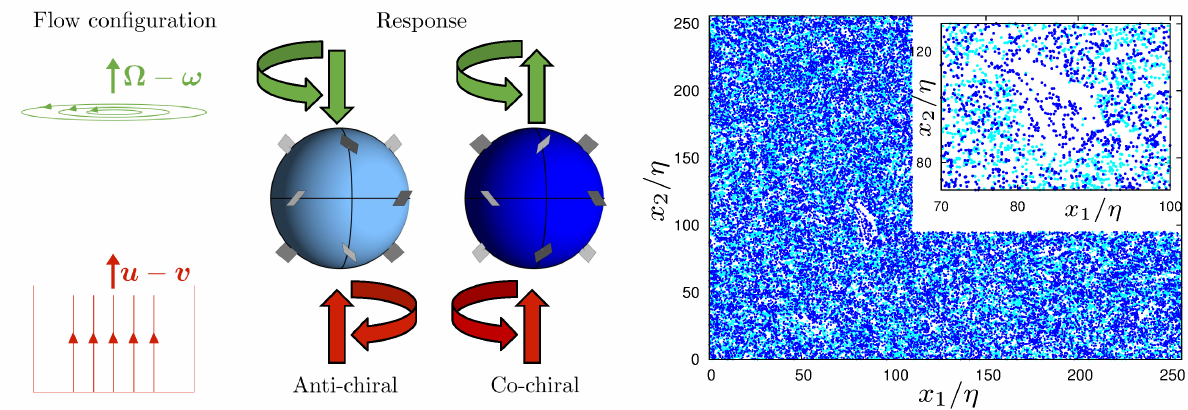}
\end{overpic}
\caption{
Left: Illustration of two isotropic helicoids with opposite helicoidality as suggested by Kelvin~\citep{Kel71}, anti-chiral ($C_0<0$, left panel) and co-chiral ($C_0>0$, right panel).
The initial response to two simple flow configurations are illustrated with arrows.
In response to an applied vertical difference in velocity, both helicoids accelerate in the direction of relative velocity, while their angular accelerations depend upon the sign of $C_0$.
Similarly, in response to an applied vertical difference in vorticity, the helicoids obtain the same angular acceleration, but they are accelerated in opposite directions.
Right: snapshot in the stationary state for two types of helicoids of opposing helicoidality in DNS of the helical turbulent flow given by the fourth case in \Tabref{param}. Points show particle positions in a slice of height $5\eta$. Parameters: $\st\approx\st_-$, $S=0.1$, $a=30$, and $C_0=-1.6$ (anti-chiral,light blue) or $C_0=1.6$ (co-chiral,blue). Inset shows a zoom to highlight that particles of different chirality accumulate in different regions.
}
 \label{fig:IsotropicHelicoids}
\end{center}
\end{figure}
One example of isotropic helicoids suggested by Kelvin~\citep{Kel71} is illustrated in \Figref{IsotropicHelicoids}.
Twelve planar vanes are attached perpendicular to the surface of a sphere at equal distances on three great circles. All vanes either form the angle $+45\degree$ (anti-chiral helicoid) or $-45\degree$ (co-chiral helicoid) with the great circle traversed clockwise, see \Figref{IsotropicHelicoids}.
The vanes cause a coupling between translational and rotational motion.
The dynamics of isotropic helicoids was studied in stationary ABC flows in~\citet{Gus16a}.
It was shown that the spatial distribution of isotropic helicoids depends on the relative chirality between the particle and the underlying flow.
We anticipate that this is also what happen in helical turbulence as  illustrated in \Figref{IsotropicHelicoids} where we show that isotropic helicoids move to different flow regions depending on their helicoidality also in our DNS of the NSE (\ref{eq:nse}).
In this section we use DNS, the stochastic model and theoretical approaches to analyse the motion of isotropic helicoids in helical turbulence.
\subsection{Isotropic helicoids}
\label{sec:eqm}
\noindent
The dynamics of an isotropic helicoid with position $\ve x$, velocity $\ve v$ and angular velocity $\ve \omega$ suspended in a fluid with  velocity $\ve u$ and vorticity $2\ve\Omega=\ve\nabla \times \ve u$ is governed by the following equations~\citep{Kel71,Hap12,Gus16a}
\begin{align}
\begin{split}
\dot{{\ve v}}= &\frac{1}{\tau_{\rm p}}\left[ {\ve u}({\ve x}(t), t) - {\ve v} + \frac{2\tilde{a}}{9}C_0 ({\ve \Omega}({\ve x}(t), t) - {\ve \omega}) \right]\\
\dot{{\ve \omega}}= &\frac{1}{\tau_{\rm p}}\left[ \frac{10}{3}S({\ve \Omega}({\ve x}(t), t) - {\ve \omega}) + \frac{5}{9\tilde{a}}C_0 ({\ve u}({\ve x}(t), t) - {\ve v}) \right]\,.
\end{split}
\label{eq:eqm_isotropic_helicoid}
\end{align}
Here dots denote time derivatives and ${\ve u}$ and ${\ve \Omega}$ are evaluated at the particle position ${\ve x}(t)$.
The dynamics of isotropic helicoids couples individual vector components of translational and rotational motion, but does not mix different components.
The dynamics is governed by four parameters.
First, $\tau_{\rm p}$ is a relaxation time quantifying particle inertia.
In the limit of $\tau_{\rm p}\to 0$ the particle approaches the dynamics of a tracer, $\ve v=\ve u$ and $\ve\omega=\ve\Omega$.
Second, $\tilde{a} = \sqrt{5I_0/2m}$ is a measure of the particle size defined by its mass $m$ and moment of inertia $I_0$.
Third, $C_0$ is the helicoidality. It quantifies the strength of the coupling between translational and rotational degrees of freedoms.
Finally, $S$ is the structural number that quantifies how much the rotational inertia of the isotropic helicoid differs from that of a spherical particle.
When $C_0=0$, the particle dynamics is that of an isotropic particle, and if further $S=1$, the dynamics is that of a spherical particle with Stokes relaxation time $\tau_{\rm p}$.
When $C_0\ne 0$, invariance of the particle dynamics under mirror reflections of the particle is broken.
Depending on the relative sign between $C_0$ and components of $\ve\Omega$, the particle accelerates either along the vorticity component, or opposite to it, see \Figref{IsotropicHelicoids}.
The only constraint on the parameters is $|C_0|<\sqrt{27S}$, required for the kinetic energy of the particle to remain finite.
The actual size of the particle, $\sim\tilde{a}$, should also be less than the smooth scale of the flow (a multiple of the Kolmogorov length $\eta$) for the point-particle approximation to be valid.
\noindent
The governing equations \eqnref{eqm_isotropic_helicoid} exemplify why isotropic helicoids are simpler extensions to spherical particles than spheroids: the dynamics of spheroids depends on their instantaneous direction in addition to $\ve v$ and $\ve\omega$ and it couples different components of the velocity and angular velocity. Moreover, in the limit of inertialess spheroids, the particle angular velocity does not simply follow $\ve\Omega$, but is also affected by the strain rate of the flow~\citep{Jef22}.

\noindent
Rescaling to dimensionless units $t'=t/\tau_\eta$, $\ve x'=\ve x/\eta$, ${{\ve u}}' = {\ve u}\tau_\eta/\eta$,
${{\ve v}}' = {\ve v}\tau_\eta/\eta$, ${{\ve \omega}}' = {\ve \omega}\tau_\eta$, and ${{\ve \Omega}}' = {\ve \Omega}\tau_\eta$ and dropping the primes in what follows,
we can write the equations of motion for each pair of components $v_i$ and $\omega_i$ in dimensionless form:
\begin{equation}
\begin{pmatrix}
\dot{v}_i\cr
\dot{\omega}_i
\end{pmatrix}
=\ma D
\begin{pmatrix}
u_i-v_i\cr
\Omega_i-\omega_i
\end{pmatrix}
\,,\hspace{0.5cm}
\ma D=
\frac{1}{\st}\begin{pmatrix}
1 & \frac{2 C_0 a}{9} \cr
\frac{5 C_0}{9 a} & \frac{10}{3}S
\end{pmatrix}\,.
\label{eq:eqm_isotropic_helicoid_matrix}
\end{equation}
Here we have introduced the dimensionless size $a={\tilde a}/\eta$ and the Stokes number $\st=\tau_{\rm p}/\tau_\eta$.
Interpreting the two-tensor $\ma D$ as a matrix, it has two eigenvalues $d_\pm$ and corresponding eigenvectors $\ve\xi_\pm$ given by
\begin{align}
\label{eq:d_pm}
d_\pm&=\frac{1}{18\st}\left( 9 + 30S \pm\sqrt{40C_0^2+9(3-10S)^2}\right) \equiv \frac{\st_{\pm}}{\st}\\
\ve\xi_\pm &=
\frac{1}{\sqrt{(2C_0a)^2+81(\st_\pm-1)^2}}
\begin{pmatrix}
2C_0a \cr
9(\st_\pm-1)
\end{pmatrix}\,.
\label{eq:xi_pm}
\end{align}
These equations are well defined for all parameter values, but in the limit of isotropic particles, $C_0\to 0$, there is a complication.
Taking the limit $C_0\to 0$ in Eqs. \eqnref{d_pm} and \eqnref{xi_pm} we need to distinguish the two cases of $S<3/10$ and $S>3/10$, resulting in the eigenvalues and eigenvectors given in \Tabref{eigensystem}.
\begin{table}
\begin{center}
\begin{tabular}{lcccc}
Condition & $\st_-$ & $\st_+$ & $\xi_-$ & $\xi_+$\\[3pt]
$S>3/10$ & $1$ & $10S/3$ & $(1, 0)$ & $(0, 1)$\\
$0<S<3/10$ & $10S/3$ & $1$ & $(0, -1)$ & $(1, 0)$
\end{tabular}
\caption{\label{tab:eigensystem}Rescaled eigenvalues $\st_\pm=d_\pm\st$ and corresponding eigenvectors $\xi_\pm$ in Eqs. \eqnref{d_pm} and \eqnref{xi_pm} for isotropic particles, $C_0=0$.}
\end{center}
\end{table}
The reason is that when $C_0=0$, the eigenvalues cross at $S=3/10$, meaning that the translational eigenvalue switches from being the largest ($d_+$) when $S<3/10$ to the smallest ($d_-$) when $S>3/10$.
Moreover, the translational and rotational degrees of freedom decouple when $C_0=0$ and the eigenvector $(1,0)$ corresponding to the translational dynamics must be $\xi_+$ for $S<3/10$ and $\xi_-$ for $S>3/10$ with a discontinuous jump at $S=3/10$.
In the same way, the eigensystem corresponding to the rotational dynamics has a discontinuity at $S=3/10$.
When $C_0=0$ the translational dynamics has a single scale of inertia. When $S\sim 1$ this scale is $\st\sim 1$, see \Tabref{eigensystem}.
This has been observed in simulations of inertial particles, where the most interesting dynamics occurs around values of $\st$ of order unity, see for example~\citet{Fes94,Bec07,Fal07}.
The translational dynamics of isotropic helicoids on the other hand has two characteristic inertial scales $\st_-$ and $\st_+$ that depend on the helicoid parameters $C_0$ and $S$. These scales may be well separated in the meaning that $\st_+/\st_-$ can take arbitrarily large values.
We therefore expect that isotropic helicoids may show significantly different behaviour depending on whether the Stokes number $\st$ is of the order of $\st_-$ or $\st_+$.
Below we illustrate this by numerical simulations and analysis of two different limiting cases. We remark that all statistical measurements have been made after that the particle dynamics and the flow velocity reached stationarity. Moreover, all considered statistical quantities are related to clustering in sub-viscous scales where we expect weak dependence on the Reynolds number~\citep{Bec07}.

\subsection{Preferential sampling of vorticity and helicity}
\label{sec:preferential_sampling}
\noindent
Inertial spherical particles are subjected to preferential sampling of particular flow structures as well as small-scale fractal clustering \citep{Max87,Fes94,Bec03,Gus16b}.
In the limit of small Stokes numbers the mechanism for clustering can be explicitly related to preferential sampling.
In~\citet{Gus16a} the  divergence of the velocity field along the trajectory of an isotropic helicoid was derived for small values of $\st$ ($\st\ll\st_-$)
\begin{equation}
\label{eq:compressibility}
\ve\nabla\ccdot\ve v \sim -\,\frac{\st}{27S-C_0^2}\Big(27S \tr[\ma A^2] -\frac{9a C_0}{5} \tr[\ma A \ma V]\Big)+\textit{o}(\st)\,,
\end{equation}
where {$\ma A$ and $\ma V$ are matrices with elements} $A_{ij}=\partial_j u_i$ and $V_{ij} = \partial_j \Omega_i$.
Depending on the sign of $\ve\nabla\ccdot\ve v$  trajectories of close-by particles may either converge ($\ve\nabla\ccdot\ve v<0$) or diverge ($\ve\nabla\ccdot\ve v>0$).
It is expected that particles cluster in regions where $\ve\nabla\ccdot\ve v<0$, i.e. where $27S \tr[\ma A^2]>9a C_0 \tr[\ma A \ma V]/5$.
For heavy spherical particles $C_0$ is zero and particles cluster in straining regions of the flow where $\tr[\ma A^2]>0$~\citep{Max87}.
For helicoids the structures in which particles with small values of $\st$ converge are more intricate and depend {in addition} on the particle parameters, $a$, $C_0$, $S$, combined with the local flow helicity as expressed in the last term, {$\propto\tr[\ma A \ma V]$,} on the right-hand side of Eq.~(\ref{eq:compressibility}).
As observed by~\citet{Gus16a}, for a flow region with strong helical coherence, $\ma V\sim c\ma A$, particles cluster where $(27S -9a cC_0 /5)\tr[\ma A^2]$ is positive.
As a consequence, particles of opposite helicoidality (different signs of $C_0$) may accumulate in flow regions of opposite sign of helicity $c$. As a result, even if the helicoids are heavier than the surrounding flow, they may cluster in vortical regions where  $\tr[\ma A^2]<0$, similar to light spherical particles.
\noindent
In order to quantify the preferential sampling of helical flow structures, we simulate the dynamics \eqnref{eqm_isotropic_helicoid_matrix} for a number of parameters summarized in \Tabref{parameters} using the flows described in Sec. \ref{sec:numerical}.
For each set of parameters, once the fluid reaches its statistically stationary state it is seeded with $2.4\times10^6$ particles.
The initial velocity and angular velocity of each particle are given by the fluid velocity and half the fluid vorticity evaluated at the particle position.
\begin{table}
\begin{center}
\begin{tabular}{lcccccc}
Helicoids type & $C_0$ & $S$ & $a$ & $\st_-$ & $\st_+$ & $\beta_{\rm eff}$ \\[3pt]
Anti-chiral & $-5$ & $1$ & $10$ & $0.058$ & $4.3$ & $-0.1$\\
 & $-1.6$ & $0.1$ & $30$ & $0.013$ & $1.3$ & $-0.2$\\
Neutral   & $0$ & $1$ & $-$ & $1$ & $3.3$ & $0$\\
Co-chiral & $1.6$ & $0.1$ & $30$ & $0.013$ & $1.3$ & $1.7$\\
 & $5$ & $1$ & $10$ & $0.058$ & $4.3$ & $0.5$
\end{tabular}
\caption{\label{tab:parameters}Overview of the five parameter families in the simulations in \Figref{Isotropic_Helicoids_Delta_H}.
For each family, $\st$ varies over a few decades.
The dynamics of the helicoids is driven by the helicoidality $C_0$, shape factor $S$, and particle size $a = {\tilde a}/\eta$ which define the characteristic scales $\st_-$, $\st_+$ and $\beta_{\rm eff}$ as introduced in Eqs. \eqnref{d_pm} and \eqnref{beta_eff} respectively; for details see \Secref{eqm}. The notion of being co-chiral or anti-chiral is made in terms of the flow helicity which is always taken in average positive in this paper. The parameter $c$ entering in the definition of $\beta_{\rm eff}$ is obtained as $c=(\langle\ve\Omega^2\rangle/\langle\ve u^2\rangle)^{1/2}=0.1$ in DNS.}
\end{center}
\end{table}
\noindent
\Figref{IsotropicHelicoids} illustrates that isotropic helicoids of opposing chirality preferentially sample different flow regions in DNS of a helical turbulent flow.
\Figref{Isotropic_Helicoids_Delta_H}{\bf a},{\bf b} shows the fraction of particles in rotational regions of the flow.
The data are plotted against $\st/\st_-$ in \Figref{Isotropic_Helicoids_Delta_H}{\bf a} and $\st/\st_+$ in \Figref{Isotropic_Helicoids_Delta_H}{\bf b}, i.e. against the inverse of the two eigenvalues \eqnref{d_pm} of the dynamics (when $C_0=0$ the data are only plotted against $\st/\st_-$ because for this case the preferential sampling cannot depend on $\st_+$).
In rotational regions of a flow, the fluid gradient matrix $\ma A$ has complex eigenvalues, or equivalently, the sign of the discriminant
\begin{equation}
\Delta = \left( \frac{{\rm det}\ma A}{2} \right)^2 - \left( \frac{{\rm tr}[\ma A^2]}{6} \right)^3\,,
\label{eq:discri}
\end{equation}
is positive~\citep{Cho90}.
We observe that isotropic helicoids with the same helicoidality (positive) of the underlying flow, $C_0=1.6$ (filled blue boxes), depend intricately on the Stokes number: for small values of $\st$ they behave similar to light particles that oversample rotational regions where $\Delta>0$, while for larger values of $\st$ they instead behave as heavy inertial particles that oversample strain regions where $\Delta<0$.
In contrast, for the other considered values of $C_0$, the helicoids always behave as heavy particles and oversample strain regions to different degrees depending on the particle parameters.
\begin{figure}
\begin{center}
\begin{overpic}[width=13.8cm,draft=false]{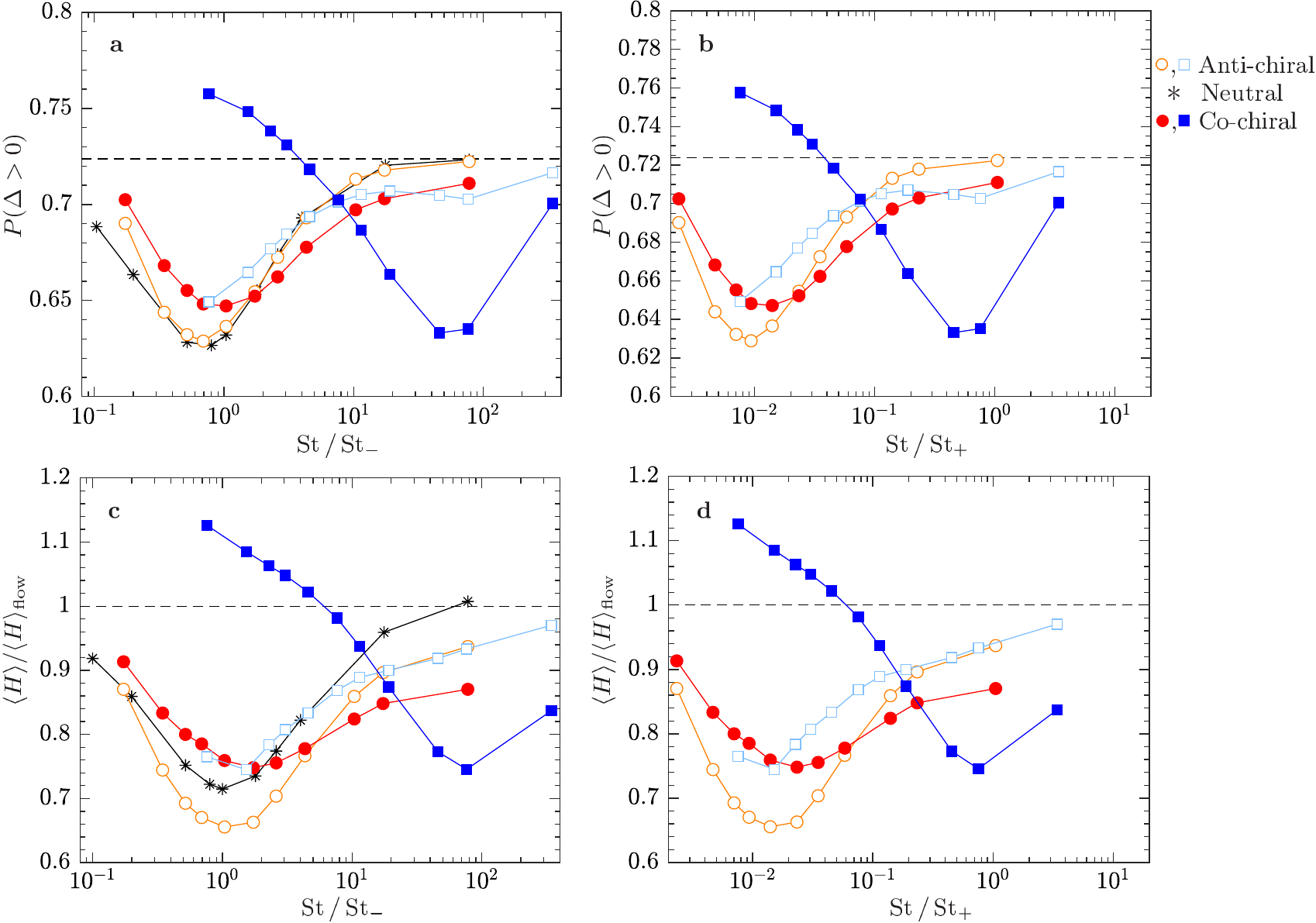}
\end{overpic}
\caption{
Upper row: fraction of particles in rotational flow regions ($\Delta > 0$) as functions of {\bf a} $\st/\st_-$ and {\bf b} $\st/\st_+$ for the DNS given by the fourth case in \Tabref{param}.
Lower row: mean fluid helicity $\langle H\rangle=2\langle{\ve u} \ccdot {\ve \Omega}\rangle$ along particle trajectories as functions of {\bf c} $\st/\st_-$ and {\bf d} $\st/\st_+$ for the DNS. The data is normalized by the helicity of the flow, $\langle H\rangle_{\rm flow}$ which is chosen to be positive in all our simulations.
The parameters of the simulations are given in \Tabref{parameters} and the simulation results are displayed as interconnected markers.
Results for neutral particles, $C_0=0$, are shown as black asterisks.
Results for $S=1$, $a=10$ helicoids are shown as hollow orange circles (anti-chiral, $C_0=-5$) and filled red circles (co-chiral, $C_0=5$).
Results for $S=0.1$, $a=30$ helicoids are shown as hollow light blue boxes (anti-chiral, $C_0=-1.6$) and filled blue boxes (co-chiral, $C_0=1.6$).
Black dashed lines show $P(\Delta>0)$ and $\langle H\rangle$ for tracer particles.
}
\figlab{Isotropic_Helicoids_Delta_H}
\end{center}
\end{figure}
\noindent
\Figref{Isotropic_Helicoids_Delta_H}{\ve c},{\bf d} shows the mean value of fluid helicity evaluated along particle trajectories,
$\langle H({\ve x(t)})\rangle=2\langle{\ve u} \ccdot {\ve \Omega}\rangle$, as functions of $\st/\st_-$ (\Figref{Isotropic_Helicoids_Delta_H}{\ve c}) and $\st/\st_+$ (\Figref{Isotropic_Helicoids_Delta_H}{\ve d}).
Comparing \Figref{Isotropic_Helicoids_Delta_H}{\bf a} and~{\bf c} shows that the behaviour is quite similar: helicoids with $C_0=1.6$ oversample rotational regions and have larger helicity than the underlying flow if the Stokes number is small enough.
This is consistent with these particles spending long time in rotational regions of the flow where helicity is high and mainly of a given sign due to the helical nature of the underlying flow.
Particles with the other investigated parameter values on the other hand, experience a fluid helicity that is lower than that of tracer particles.
This is consistent with these particles aggregating in fluid strain regions where helicity is small.
We also observe a transition at intermediate Stokes numbers: for small values of $\st$, isotropic helicoids with negative values of $C_0$ are more likely to sample flow regions {with low degree of helicity}, while for large values of $\st$ helicoids with positive values of $C_0$ on average sample {lower degree of helicity}.

\subsection{Small-scale fractal clustering}
\begin{figure}
\begin{center}
\begin{overpic}[width=13.8cm,draft=false]{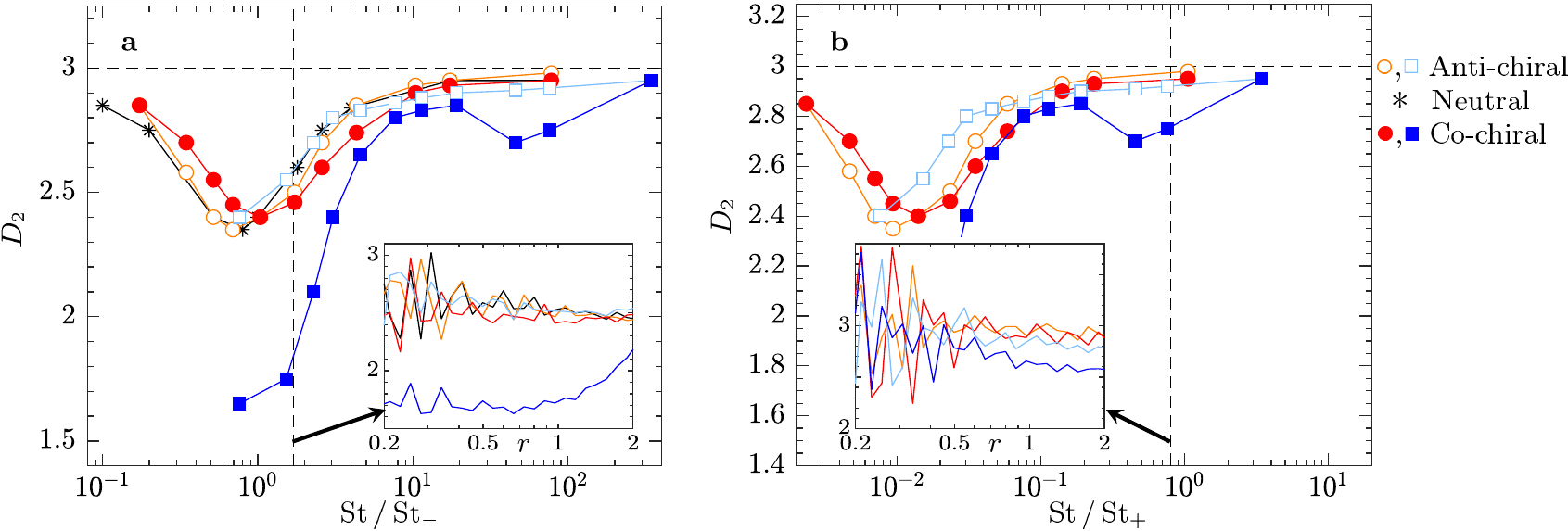}
\end{overpic}
\caption{Spatial correlation dimension $D_2$ as functions of {\bf a} $\st/\st_-$ and {\bf b} $\st/\st_+$ for the DNS and parameters of \Figref{Isotropic_Helicoids_Delta_H}. Insets show local slope $ d \log(P(r))/d \log(r)$.
}
\label{fig:5}
\end{center}
\end{figure}
\noindent
The previous section shows that helicoids of different chirality may go to very different regions in the flow.
This is exemplified in \Figref{IsotropicHelicoids} which shows that the helicoids of opposite chirality distribute in different flow regions.
It is also visible in \Figref{IsotropicHelicoids} that the spatial clustering is different in nature, close-by helicoids with $C_0=-1.6$ seem to distribute on different kinds of structures compared to helicoids with $C_0=1.6$.
It is known that spherical particles show small-scale fractal clustering due to preferential sampling and due to the dissipative nature of the dynamics.
We {investigate} the degree of spatial clustering of helicoids by computing the spatial correlation dimension, $D_2$, which defines the probability distribution $P(r)$ to find two helicoids within a small  spatial distance $r$:
$$P(r)\sim r^{D_2},\qquad r \ll 1.$$
\Figref{5}{\bf a},{\bf b} shows the behaviours of $D_2$ as functions of {\bf a} $\st/\st_-$ and {\bf b} $\st/\st_+$ for {DNS with the same parameters as in \Figref{Isotropic_Helicoids_Delta_H}}.
\noindent
Starting at small $\st/\st_-$, the correlation dimension is close to the spatial dimension and clustering is weak.
As $\st/\st_-$ is increased, the clustering increases until it reaches a maximum around $\st/\st_-\sim 1$.
Finally, as $\st/\st_-$ is further increased, the clustering becomes weaker and saturates at the spatial dimension for large values of $\st/\st_-$.
The helicoids with $C_0=1.6$ show stronger clustering around $\st/\st_-\sim 1$ compared to the other cases.

\subsection{Limiting cases}
\label{sec:limiting}
\noindent
We now explain the main features of the data shown in Figs. \figref{Isotropic_Helicoids_Delta_H} and \figref{5} by analysing three limiting cases of the system parameters.
The first two limiting cases, $\st\ll\st_+$ and $\st\gg\st_-$, apply to both DNS and the stochastic model, while the third limit is obtained for small values of $\ku$ and therefore only applies to the stochastic model.

\noindent We use Eqs. \eqnref{d_pm} and \eqnref{xi_pm} to write the dynamics \eqnref{eqm_isotropic_helicoid_matrix} in its diagonal basis as
\begin{align}
\label{eq:diag1}
\dot{\zeta}_{-,i}&=\frac{\st_-}{\st}(u_{-,i}-\zeta_{-,i})\\
\dot{\zeta}_{+,i}&=\frac{\st_+}{\st}(u_{+,i}-\zeta_{+,i})\,.
\label{eq:diag2}
\end{align}
Here we have changed basis for each component $i$ using
\begin{align}
\begin{pmatrix}
v_i\cr
\omega_i
\end{pmatrix}
=
\ma X
\begin{pmatrix}
\zeta_{-,i}\cr
\zeta_{+,i}
\end{pmatrix}
\hspace{0.5cm}\mbox{and}\hspace{0.5cm}
\begin{pmatrix}
u_i\cr
\Omega_i
\end{pmatrix}
=
\ma X
\begin{pmatrix}
u_{-,i}\cr
u_{+,i}
\end{pmatrix}
\,,
\end{align}
where the columns of the $2\times 2$ matrix $\ma X$ consist of the eigenvectors $\ve\xi_-$ and $\ve\xi_+$.

\noindent
The dimensionless parameter groups in Eqs. \eqnref{diag1} and \eqnref{diag2} can be viewed as ratios of the time scales, $\tau_\pm/\tau_\eta\equiv \st/\st_\pm$, where $\tau_\eta$ is the Kolmogorov time of the flow and $\tau_\pm$ are two particle time scales ($\tau_+\le\tau_-$) of the isotropic helicoid ($\tau_+=\tau_-=\tau_{\rm p}$ for a spherical particle).
Eqs. \eqnref{diag1} and \eqnref{diag2} are only implicitly coupled through the trajectory dependence in $\ve u_-$ and $\ve u_+$.
We therefore expect that the two limiting cases $\st\ll\st_+$ and $\st\gg\st_-$ can be taken, to a lowest-order approximation, in one of Eqs. \eqnref{diag1} and \eqnref{diag2} independent from the second equation.
In summary, when $\st\ll\st_+$ ($\tau_+\ll\tau_\eta$), \Eqnref{diag2} becomes overdamped and the remaining equation \eqnref{diag1} gives rise to strong preferential sampling when $\st/\st_-\sim 1$ ($\tau_-\sim\tau_\eta$) in analogue to the case $\st\sim 1$ ($\tau_{\rm p}\sim\tau_\eta$) for inertial spherical particles.
In the second limit $\st\gg\st_-$ ($\tau_-\gg\tau_\eta$), \Eqnref{diag1} becomes underdamped and $\zeta_-$ can to lowest order be approximated by its mean value.
The remaining equation \eqnref{diag2} gives rise to strong preferential sampling when $\st/\st_+\sim 1$ ($\tau_+\sim\tau_\eta$).
Below we discuss the two limiting cases in more detail.

\subsubsection{Case $\st\ll\st_+$}
\label{sec:limitingI}
\noindent Consider first $\st\ll\st_+$ with general values of $\st_-$.
For the acceleration to remain finite in \Eqnref{diag2} in this limit, we must have $0\sim u_{+,i}-\zeta_{+,i}$.
In terms of the original coordinates, this condition gives the following constraint:
\begin{equation}
\label{eq:constrain}
{\ve \omega}-{\ve \Omega}=\frac{9}{2C_0a}(1-\st_-)({\ve u}-{\ve v})\,.
\end{equation}
Using Eq.(\ref{eq:constrain}) and its time derivative to replace ${\ve \omega}$ and $\dot{\ve \omega}$ in \Eqnref{diag1}, and reverting to the original coordinates, we obtain:
\begin{equation}
\label{eq:caseone_mot}
\dot{{\ve v}}=\frac{\st_-}{\st}({\ve u}-{\ve v}) + \frac{1}{3(\st_- - \st_+)}\left[3(\st_- -1){\dot{\ve u}}-\frac{2C_0a}{3}{\dot{\ve \Omega}}\right]\,.
\end{equation}
Thus, a single equation determines the velocity of the isotropic helicoid in the limit $\st\ll\st_+$ and the angular velocity is given by \Eqnref{constrain}.
It can be noted that for the case of $C_0=0$, the constraint \eqnref{constrain} becomes singular because velocity and angular velocity are uncoupled.
However, \Eqnref{caseone_mot} still shows the same results as those obtained by letting $C_0=0$ and $\st\ll\st_+$ in the original equation \eqnref{eqm_isotropic_helicoid_matrix}.
When $C_0=0$ and $S>3/10$, \Eqnref{caseone_mot} simplifies to $\dot{\ve v}=(\ve u-\ve v)/\st$ while the rotational dynamics is overdamped, $\ve\omega-\ve\Omega=0$.
When $C_0=0$ and $S<3/10$ on the other hand, the condition $\st\ll\st_+=1$ implies that the translational dynamics is overdamped and \Eqnref{caseone_mot} simplifies to $\dot{\ve v}=(\ve u-\ve v)/\st+\dot{\ve u}$. This equation relaxes to the overdamped limit $\ve v=\ve u$ after a short initial transient on the time scale of order $\st\ll 1$.

\noindent
The dynamics \eqnref{caseone_mot} can be further simplified using one or both of the following two assumptions.
First, for small enough values of $\st / \st_-$, we {approximate $\dot{\ve u} \sim {\rm D}_t{\ve u}$ and $\dot{\ve\Omega} \sim {\rm D}_t{\ve \Omega}$, where ${\rm D}_t \equiv \partial_t+(\ve u\ccdot\ve\nabla)$ are advective derivatives.}
Second, in a helical flow $\ve\Omega$ and $\ve u$ tend to be aligned and we may approximate ${\ve \Omega}\sim c{\ve u}$ with some proportionality constant $c$.
Using these approximations, Eq.(\ref{eq:caseone_mot}) simplifies to:
\begin{equation}
\label{eq:caseone_mot_simple}
\dot{{\ve v}}=\frac{St_-}{St}({\ve u}-{\ve v}) + \frac{9(St_- -1)-2cC_0a}{9(St_- - St_+)}{\rm D}_t\ve u\,.
\end{equation}
If we define an effective Stokes number $\st_{\rm eff}$ and an effective density parameter $\beta_{\rm eff}$:
\begin{align}
\eqnlab{St_eff}
\st_{\rm eff}&=\frac{\st}{\st_-}\\
\beta_{\rm eff}&=\frac{9(\st_- -1)-2cC_0a}{9(\st_- - \st_+)}\,,
\eqnlab{beta_eff}
\end{align}
then \Eqnref{caseone_mot_simple} becomes identical to the equation of motion of small spherical particles~\citep{Max83} with sub-dominant terms neglected:
\begin{equation}
\dot{{\ve v}}=\frac{1}{St}({\ve u}-{\ve v}) + \beta{\rm D}_t\ve u\,.
\end{equation}
For spherical particles the density is characterized by $\beta$: $0\le\beta<1$ corresponds to heavy particles more dense than the fluid, $\beta=1$ corresponds to neutrally buoyant particles and $1<\beta\le 3$ corresponds to particles lighter than the fluid.
We remark that $\beta_{\rm eff}$ in \Eqnref{beta_eff} is not constrained to the interval $0\le\beta\le 3$ as the case of spherical particles, $\beta_{\rm eff}$ may also take negative values as well as values larger than $3$.
\\\\
{\noindent\it Comparison to DNS}\\
Although being a crude first-order approximation, \Eqnref{caseone_mot_simple} allows us to use what is known from spherical particles to explain the main features of the behaviour of helicoids in DNS for $\st\ll\st_+$  in \Figref{Isotropic_Helicoids_Delta_H}{\bf a}.
The values of the effective density parameter $\beta_{\rm eff}$ in \Eqnref{beta_eff} are quoted in \Tabref{parameters} for our parameters.
Only the case $C_0=1.6$ has $\beta_{\rm eff}$ larger than one, corresponding to spherical particles lighter than the fluid.
Such particles are expected to preferentially sample rotational flow regions when the effective Stokes number $\st_{\rm eff}=\st/\st_-$ is of order unity, which is consistent with the data in \Figref{Isotropic_Helicoids_Delta_H}{\bf a}.
The cases of helicoids with $C_0=-1.6$ or $C_0=-5$ can be viewed as heavy particles because $\beta_{\rm eff}$ is close to zero.
In these cases the helicoids preferentially sample strain regions of the fluid with strongest effect around $\st_{\rm eff} \sim O(1)$.
Finally, the case $C_0=5$ has $\beta_{\rm eff}\approx 0.5$, making it heavy but not as heavy as the cases with negative values of $C_0$.
This is consistent with a maximal preferential sampling of straining regions around $\st_{\rm eff} \sim O(1)$ that is somewhat lower than for the case $C_0=-5$, but inconsistent for the case of $C_0=-1.6$ where the maximal preferential sampling is of the same order, see \Figref{Isotropic_Helicoids_Delta_H}{\bf a}.

\noindent The approximation \eqnref{caseone_mot_simple} allows us to also explain the observed clustering in \Figref{5}{\bf a}.
The helicoids with $C_0=1.6$ show stronger clustering around $\st/\st_-\sim 1$ compared to the other cases.
This can be explained by the observation that the dynamics of helicoids with $C_0=1.6$ is similar to that of light spherical particles and that the other types of helicoids have dynamics similar to that of heavy spherical particles with effective Stokes numbers $\st/\st_-$.
Light spherical particles cluster in rotational regions of the flow and show more clustering than heavy spherical particles~\citep{Bec03,Tos09}.
This explains why the helicoids with $C_0=1.6$ have a smaller fractal dimension, close to $D_2=1.6$, than the other, effectively heavy spherical particles.
It also explains why the correlation dimension for the helicoids with $C_0=-1.6$ and $C_0=\pm5$ approximately collapse on the correlation dimension of spherical particles when plotted against $\st/\st_-$.
\\\\
{\noindent\it Comparison to stochastic model}\\
Below we study in detail the validity of the approximations leading to \eqnref{caseone_mot_simple}.
Since DNS is slow, it is hard to reach the steady state for helicoids with $\st\ll\st_-$ and to get good statistics in this limit.
We therefore use the stochastic model to study the approximations.

\begin{figure}
\begin{center}
\begin{overpic}[width=13.8cm,draft=false]{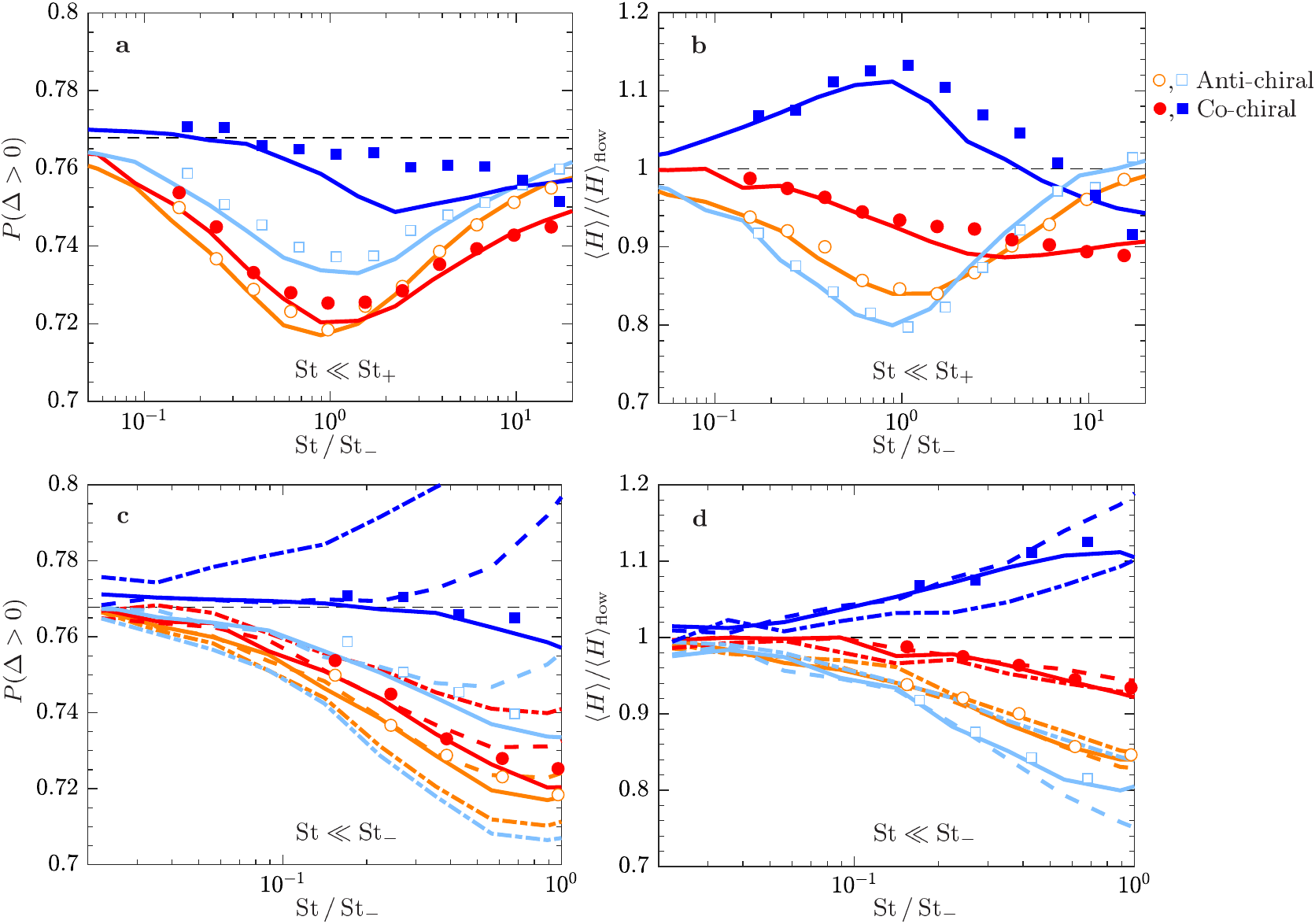}
\end{overpic}
\caption{Comparison of stochastic model simulations for ({\bf a}) the probability to be in a rotational region, $P(\Delta>0)$, and ({\bf b}) the average helicity, $\langle H\rangle$, to results in the limiting case $\st\ll\st_+$ discussed in \Secref{limitingI}.
Markers show results from simulations of the full dynamics \eqnref{eqm_isotropic_helicoid} and solid lines show the approximation \eqnref{caseone_mot} evaluated using stochastic model simulations. The upper bound $\st/\st_-\sim 20$ corresponds to $\st\sim 0.2\st_+$.
{\bf c}, {\bf d}: an enlargement of the region where  $\st\ll\st_-\ll\st_+$.
Solid lines and markers as in panels {\bf a} and {\bf b}.
Dashed lines show the approximation \eqnref{caseone_mot} with $\dot{\ve u}$ and $\dot{\ve\Omega}$ replaced by ${\rm D}{\ve u}/{\rm D}t$ and ${\rm D}{\ve \Omega}/{\rm D}t$.
Dash-dotted lines show the approximation \eqnref{caseone_mot_simple}.
Helicoid parameters corresponding to \Tabref{parameters} and \Figref{Isotropic_Helicoids_Delta_H}: $S=1$, $a=10$, hollow orange circles (anti-chiral, $C_0=-5$) and filled red circles (co-chiral, $C_0=5$).
$S=0.1$, $a=30$ hollow light blue boxes (anti-chiral, $C_0=-1.6$) and filled blue boxes (co-chiral, $C_0=1.6$).
Black dashed lines show results for tracer particles.
}
 \label{fig:4}
\end{center}
\end{figure}

\noindent\Figref{4}{\bf a},{\bf b} shows simulation results for isotropic helicoids with $\st\ll\st_+$ in the stochastic model described in \Secref{stochastic} (solid lines).
We choose parameters that are expected to correspond well with the DNS parameters in \Figref{Isotropic_Helicoids_Delta_H}.
We use a large Kubo number, $\ku=10$, corresponding to the persistent flow limit where the dynamics most resembles the small scales in turbulence.
As described in \Secref{stochastic} we fix $H_0=0.85$ to match the distribution of flow helicity to that of the DNS and we take $\eta_0/\eta=10$ to compensate for the difference between the smooth length scale of the dissipation range and the Kolmogorov length in DNS.
Finally, similarly to the DNS we base the Stokes number in the stochastic model on the Lagrangian time scale of tracer particles, $\tau_\eta\equiv\langle\tr(\ma A\ma A\T)\rangle^{-1/2}_{\rm flow}=\eta_0/(\sqrt{5}u_0)$.
In previous studies similar schemes have resulted in qualitative agreement between stochastic model simulations and DNS for the dynamics of spherical particles, elongated particles and gyrotactic microswimmers~\citep{Gus15,Gus16b,Gus17}.
Comparing \Figref{4}{\bf a} and {\bf b} to \Figref{Isotropic_Helicoids_Delta_H}{\bf a} and {\bf c} we obtain a qualitative agreement with the DNS also for the isotropic helicoids.
The general trends as functions of $\st$ for the different values of $C_0$ agree, but the detailed values disagree in some ranges.
One example is the probability of finding helicoids with $C_0=1.6$ in rotational regions with $\Delta>0$.
Although being larger than the probability for other parameter values, it is not larger than the probability of the underlying flow as for the DNS case.
One possible explanation for this is that the life time of vortex regions is longer in DNS than in the stochastic model.

\noindent
\Figref{4}{\bf a},{\bf b} also compares stochastic model simulations of the approximation \eqnref{caseone_mot} to the full simulation data of \Eqnref{eqm_isotropic_helicoid_matrix}.
We observe a quantitative agreement of the approximation in the expected limit $\st\ll\st_+$.
\Figref{4}{\bf c},{\bf d} shows numerical simulations of \eqnref{caseone_mot} with $\dot{\ve u}$ and $\dot{\ve\Omega}$ replaced by ${\rm D}_t{\ve u}$ and ${\rm D}_t{\ve \Omega}$, which approach the results of \Eqnref{caseone_mot} when $\st/\st_-\ll 1$ as expected.
Finally, \Figref{4}{\bf c},{\bf d} also shows the approximation \eqnref{caseone_mot_simple} using ${\ve \Omega}=c{\ve u}$ with $c=(\langle\ve\Omega^2\rangle/\langle\ve u^2\rangle)^{1/2}=\sqrt{5}/20$ for the stochastic model (rescaled using $\eta_0=10\eta$).
We observe that the prediction using spherical particles reproduces the average sampling of helicity well for $\st\ll\st_-$ (\Figref{4}{\bf d}), while it does not work as well for the probability of being in vortex regions (\Figref{4}{\bf c}).
Discrepancies in this approximation are expected because the fluid velocity and vorticity are not perfectly aligned in the helical flow with $H_0=0.85$.
In particular, the approximation fails for the probability of being in vortex regions for the cases with $C_0\pm1.6$: for $C_0=1.6$ the helicoids do not oversample vortex regions to the degree that is predicted by \Eqnref{caseone_mot_simple} and for $C_0=-1.6$ the helicoids show a larger probability than is predicted.
As discussed above, this is in contrast to DNS which agrees better with the trends predicted by \Eqnref{caseone_mot_simple}.

\begin{figure}
\begin{center}
\begin{overpic}[width=13cm,draft=false]{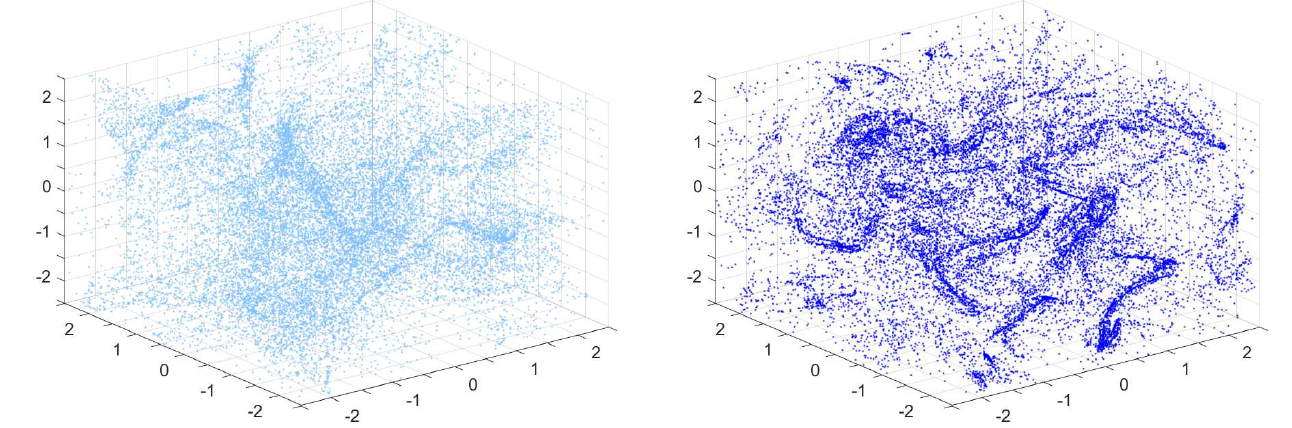}
\end{overpic}
\caption{
Snapshot of positions of isotropic helicoids in stochastic model simulations for a flow with $\ku=10$ and maximal helicity $H_0\approx 2.1$ ($\mu\to\infty$), and particle parameters $\st=0.9\st_-$, $a=30$, $S=0.1$,  and $C_0=-1.6$ (left) and $C_0=1.6$ (right).
The history of the underlying flow is identical for the two simulations.
Coordinate axes are dedimensionalized using $\eta_0$.
}
 \label{fig:positions}
\end{center}
\end{figure}

\begin{figure}
\begin{center}
\begin{overpic}[width=13cm,draft=false]{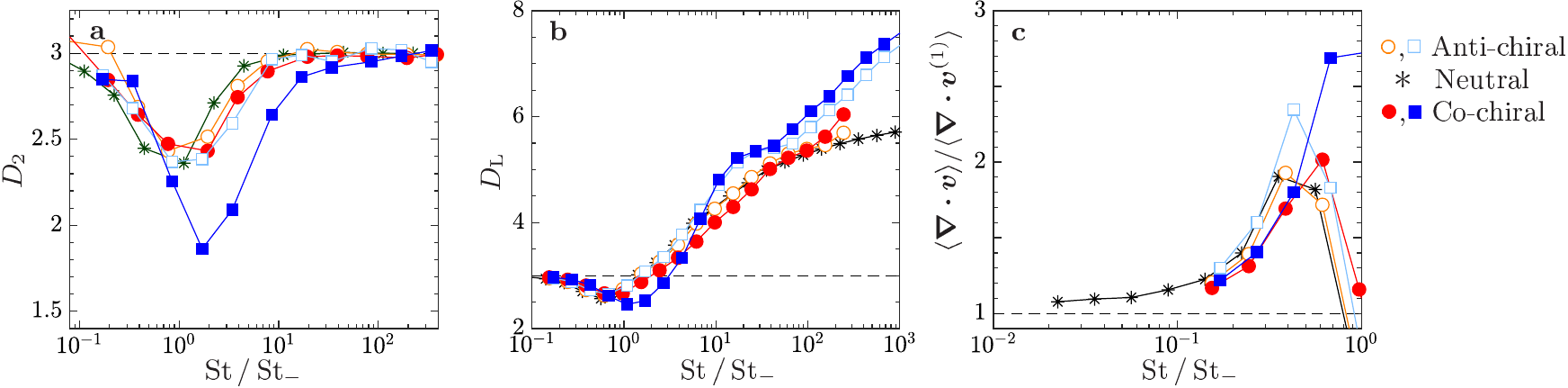}
\end{overpic}
\vspace{0.2cm}
\caption{
Stochastic model simulations of {\bf a} Correlation dimension $D_2$, {\bf b} phase-space Lyapunov dimension $D_{\rm L}$ and {\bf c} the relative amplitude of average compressibility along particle trajectories estimated using the sum of the first three Lyapunov exponents, $\langle\ve\nabla\ccdot\ve v\rangle=\lambda_1+\lambda_2+\lambda_3$ and the approximation \eqnref{compressibility}, $\langle\ve\nabla\ccdot\ve v^{(1)}\rangle$.
Parameters as in \Figref{4} with the addition of spherical particles (black asterisks).
}
 \label{fig:Nablav}
\end{center}
\end{figure}

\noindent
\Figref{positions} shows the positions of isotropic helicoids in a strongly helical flow in the stochastic model.
Similar to the DNS in \Figref{IsotropicHelicoids}, helicoids of opposing chirality go to different regions in the flow and close-by anti-chiral particles form structures of a different kind than co-chiral helicoids.
\Figref{Nablav}{\bf a} shows stochastic model results for the correlation dimension $D_2$ with parameters corresponding to the DNS in \Figref{5}.
We observe qualitative agreement between \Figref{Nablav}{\bf a} and the DNS.

Using $D_2$ to quantify the fractal dimension of clustering may be problematic because it often converges slowly and consequently very small scales must be resolved in the distribution $P(r)$.
An alternative quantification of fractal clustering which is easier to evaluate accurately is provided by the Lyapunov dimension (Kaplan-Yorke dimension) $D_{\rm L}$~\citep{Kap83}.
Denoting by $\lambda_1\ge\lambda_2\ge\dots\ge\lambda_9$ the nine Lyapunov exponents of the equation system \eqnref{eqm_isotropic_helicoid} together with $\dot{\ve x}=\ve v$, the Lyapunov dimension is
\begin{align}
D_{\rm L}=K+\sum_{i=1}^K\lambda_i/|\lambda_{K+1}|\,,
\eqnlab{DL_def}
\end{align}
where $K$ is the largest integer such that $\sum_{i=1}^K\lambda_i>0$.
\Figref{Nablav}{\bf b} shows numerical evaluation of the Lyapunov dimension for the stochastic model.
For $D_{\rm L}<3$ it shows similar trends as the correlation dimension $D_2$.
The values of $D_{\rm L}$ are somewhat higher than $D_2$, consistent with the fractal attractor in turbulence being a multifractal with $D_{\rm L}\ge D_2$~\citep{Bec05}.
Finally, we remark that the Lyapunov dimension defined in \Eqnref{DL_def} describes the dimension of the fractal in phase space, and it is therefore not bounded by the spatial dimension 3 as is the case for the spatial correlation dimension.
For large values of $\st$, the Lyapunov dimension is expected to approach the dimensionality $D$ of phase space.
This is consistent with the data in \Figref{Nablav}{\bf b} ($D=9$ for helicoids and $D=6$ for spherical particles).

\noindent
Evaluation of the Lyapunov exponents also allows us to validate the approximation \eqnref{compressibility} of the local compressibility in the stochastic model.
The long-term average of the compressibility along particle trajectories is identical to the sum over the first three Lyapunov exponents, $\langle\ve\nabla\ccdot\ve v\rangle=\lambda_1+\lambda_2+\lambda_3$.
Using this relation, we verify in \Figref{Nablav}{\bf c} the small $\st$ approximation given by \Eqnref{compressibility}.
We remark that the limit used to obtain \Eqnref{compressibility} is the same as in this section ($\st\ll\st_+$), but requires first-order corrections in $\st/\st_+$ to the condition in \Eqnref{constrain}.
Moreover, we need to consider $\st/\st_-\ll 1$.
Consistently with this limit, we
find that the {average of the} approximation \eqnref{compressibility} tends to approach {$\langle\ve\nabla\ccdot\ve v\rangle$} as $\st/\st_-$ is reduced and that the two expressions {agree} approximately for $\st/\st_-\sim0.1$.

\subsubsection{Case $\st\gg\st_-$}
\label{sec:limitingII}
\noindent We now consider the second limiting case, $\st \gg \st_-$ with general values of $\st_+$. In this limit $\zeta_{-,i}$ in \Eqnref{diag1} responds slowly to changes in the flow compared to $\zeta_{+,i}$.
Due to the symmetries of the underlying flow, we expect the averages of $u_{-,i}$ and consequently of $\zeta_{-,i}$ to vanish.
We therefore replace $\zeta_{-,i}$ by its vanishing average, $\zeta_{-,i}=0$,
which gives the following constraint on $\ve\omega$
\begin{align*}
\ve\omega=\frac{9(\st_+ - 1)}{2aC_0}\ve v\,.
\end{align*}
Inserting this constraint and its time derivative into \Eqnref{diag2}, gives the following equation for the velocity
\begin{align}
\dot{\ve v}=\frac{\st_+}{\st}\left[\frac{3(3\st_+ - 10S)\ve u + 2aC_0\ve\Omega}{9(\st_+ - \st_-)} - \ve v\right]\,.
\label{eq:casetwo}
\end{align}
As for the first limiting case we can consider a helical flow with $\ve\Omega\sim c\ve u$ to obtain
\begin{align}
\dot{\ve v}=\frac{\st_+}{\st}\left[\beta_{\rm eff}\ve u - \ve v\right]\,,
\label{eq:casetwo_simple}
\end{align}
where $\beta_{\rm eff}$ is the parameter in \Eqnref{beta_eff} occurring in the limit of $\st\ll\st_-$.
Thus, in the limit $\st \gg \st_-$ the equation of motion for helicoids in a helical flow is like a Stokes drag with effective Stokes number $\st/\st_+$ and with a rescaled amplitude of the fluid velocity.
\\\\
{\noindent\it Comparison to DNS}\\
The limiting dynamics in \Eqnref{casetwo_simple} allows us to explain the DNS results in \Figref{Isotropic_Helicoids_Delta_H}{\bf b}.
For helicoids with negative values of $C_0$, the coupling to the flow, $\beta_{\rm eff}\ve u$ in \Eqnref{casetwo_simple}, is small, see \Tabref{parameters}.
The particle motion is thus expected to be only weakly correlated to the underlying flow structures, which is consistent with the data: the particles with negative values of $C_0$ have approximately the same statistical properties as the flow (black dashed lines in \Figref{Isotropic_Helicoids_Delta_H}{\bf b}).
The helicoids with positive values of $C_0$ on the other hand have $\beta_{\rm eff}\sim 1$ and are therefore expected to have preferential sampling similar to spherical particles with effective Stokes number $\st_+/\st$.
This is what we observe for $C_0=1.6$, the shape and magnitude of the curve around $\st\sim\st_+$ is similar to that of spherical particles.
However, the approximation \eqnref{casetwo_simple} does not work as well for $C_0=5$: even though $\beta_{\rm eff}\approx 0.5$ the preferential sampling is only slightly larger than the case $C_0=-5$.
Since $C_0=5$ also only show small agreement with the predictions in the first limiting case $\st\ll\st_+$, we conclude that the approximations (for example $\ve\Omega\sim c\ve u$) leading to \Eqnref{caseone_mot_simple} and \Eqnref{casetwo_simple} may not be so accurate for $C_0=5$ in DNS.

\begin{figure}
\begin{center}
\begin{overpic}[width=13.8cm,draft=false]{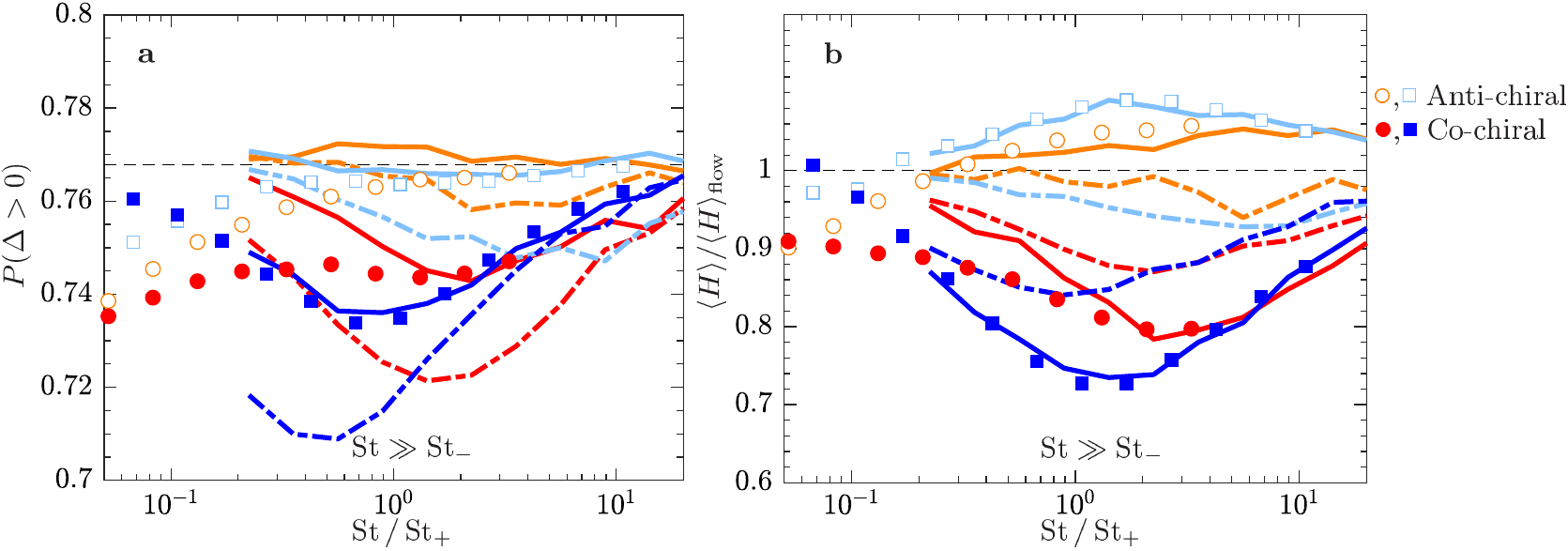}
\end{overpic}
\caption{
Comparison of stochastic model simulations for ({\bf a}) the probability to be in a rotational region, $P(\Delta>0)$, and ({\bf b}) the average helicity, $\langle H\rangle$, to results in the limiting case $\st\gg\st_-$ discussed in \Secref{limitingII}.
 Markers show results from simulations of the full dynamics \eqnref{eqm_isotropic_helicoid}, same parameters as in \Figref{4}, solid lines show the approximation \eqnref{casetwo} and dash-dotted lines show the refined approximation~\eqnref{casetwo_simple} for flows where fluid vorticity and velocity aligns. The lower bound $\st/\st_+\sim 0.05$ corresponds to $\st\sim 5\st_-$.
}
 \label{fig:4case2}
\end{center}
\end{figure}

Using the approximation \eqnref{casetwo_simple} to explain the fractal clustering observed in \Figref{5}{\bf b}, we would expect that the anti-chiral helicoids, having small coupling to the flow, should show small clustering ($D_2\approx d$) and that the co-chiral helicoids should show larger clustering of the same order as the spherical particles.
The numerical data for $C_0=1.6$ indeed show a second peak of clustering (minimum of $D_2$) around $\st\sim\st_+$.
However, we remark that the result for $D_2$ is measured at finite separation which might not reflect the true asymptotic scaling for $r \to 0$.
Indeed, as seen in the inset of \Figref{5}{\bf b}, the local slope of $D_2$ for $C_0=1.6$ seems to drift towards larger values as $r$ is decreased in the range of $r$ we can resolve.
This can be explained by the fact that deviations from the approximation \eqnref{casetwo_simple} {depend on the flow histories experienced by the particles and are therefore} different for {two close-by} particles, which results in a uniform distribution of particles for small enough scales.
{In contrast, deviations from the overdamped approximation \eqnref{caseone_mot_simple} mainly depend on the instantaneous flow and are therefore approximately the same for the two particles.}
\\\\
{\noindent\it Comparison to stochastic model}\\
\noindent\Figref{4case2} shows simulation results in the range $\st\gg\st_-$ for isotropic helicoids in the stochastic model, plotted against $\st/\st_+$.
Similar to the case $\st\ll\st_+$, comparison between the full model data in \Figref{4case2}{\bf a} and {\bf b} to the corresponding DNS data in \Figref{Isotropic_Helicoids_Delta_H}{\bf b} and {\bf d} shows similar trends as functions of $\st$, while the details differ in some ranges.
The co-chiral helicoids, having $\beta_{\rm eff}\sim 1$ show preferential sampling of straining regions with $\st\sim\st_+$, while the anti-chiral helicoids have negligible preferential sampling of straining regions.

\noindent We compare stochastic model simulations of the approximation \eqnref{casetwo} [solid lines] to simulations of \Eqnref{eqm_isotropic_helicoid} in \Figref{4case2}.
When $\st\gg\st_-$ we observe a quantitative agreement for $C_0=\pm 1.6$ and qualitative agreement when $C_0=\pm 5$.
\Figref{4case2} also shows that the approximation \eqnref{casetwo_simple} based on $\ve\Omega=c\ve u$ (dash-dotted lines) does not work equally well. \Eqnref{casetwo_simple} reproduces that co-chiral helicoids have larger preferential sampling of straining regions around $\st\sim\st_+$ than anti-chiral helicoids, but the degree of preferential sampling does not come out correctly in general.
We conclude by remarking that, as expected, in the underdamped limit of $\st\gg\st_+$ the preferential sampling for all parameter cases in \Tabref{parameters} converges to that of the flow (not shown).

\noindent As seen in \Figref{Nablav}{\bf a} the correlation dimension $D_2$ increases monotonously towards the spatial dimension $d=3$ after the peak at $\st\sim\st_-$.
As a consequence, $D_2$ does not show any clustering around $\st\sim\st_+$.
The deviations from the DNS data in \Figref{5}{\bf b} can be explained by the fact that the observed data are better resolved in the stochastic model: the correlation dimension around $\st\sim\st_+$ shows a scaling $P(r)\sim r^{D_2(r)}$ with local exponent $D_2(r)<3$ for a range of $r \ll 1$ (similar to the DNS in this range), while for small enough values of $r$, the uniform $D_2=3$ scaling is approached (not shown).

\subsubsection{Small values of $\ku$}
\begin{figure}
\begin{center}
\begin{overpic}[width=13.8cm,draft=false]{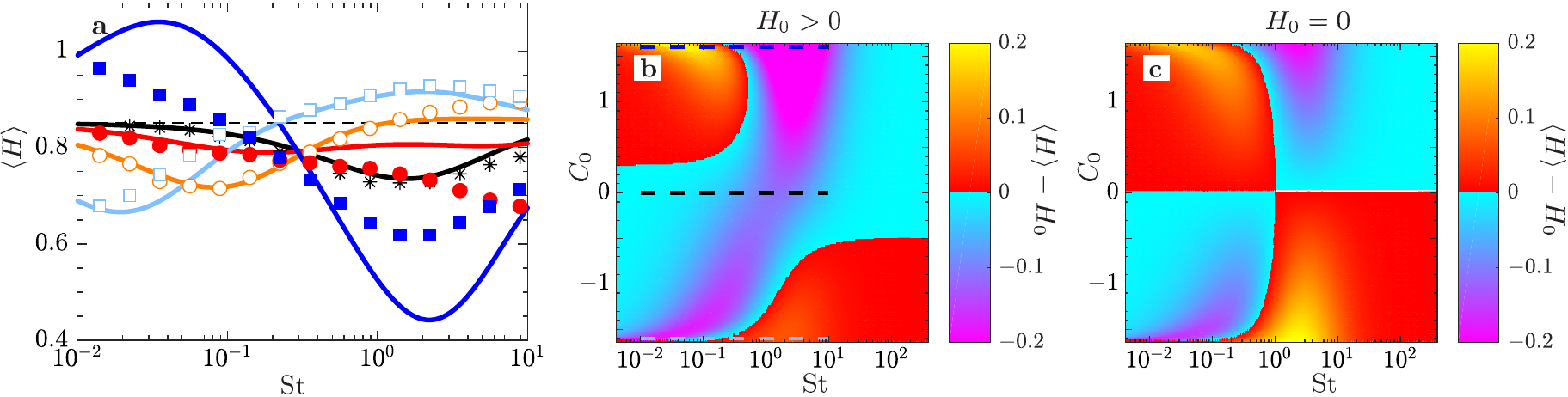}
\end{overpic}
 \caption{
 Evaluation of theory \eqnref{meanHsmallKu} for mean helicity $\langle H\rangle$ for small values of $\ku$ with rescaled parameters $\ku$ and $\st$, see text.
 {\bf a} Comparison of the theory \eqnref{meanHsmallKu} to simulation data for $\ku=10$.
 Markers correspond to the data in \Figref{4}{\bf b} (without division by $H_0$ and plotted against $\st$).
 Solid lines show \Eqnref{meanHsmallKu}. The five parameter combinations of $C_0$, $S$ and $a$ are plotted as five lines with the colors of the corresponding markers.
 {\bf b} Heat map of the theory \eqnref{meanHsmallKu} for the deviation of average helicity, $\langle H\rangle$, from that of the flow, $\langle H\rangle_{\rm flow}=H_0$, plotted against $\st$ and $C_0$ for $S=0.1$, $a=30$ and $H_0=0.85$.
 Dashed lines correspond to the curves with $C_0=-1.6$ (light blue), $C_0=0$ (black), and $C_0=1.6$ (blue) shown in panel {\bf a}.
 {\bf c} Same as {\bf b} but for a neutral flow, $H_0=0$.
}
 \label{fig:smallKu}
\end{center}
\end{figure}

\noindent
In the stochastic model the properties of the flow can be modified by changing the value of the Kubo number \eqnref{Kubo}.
In general, this allows exploration of the robustness of results with respect to the nature of the flow.
In the limit of small $\ku$ we can solve the dynamics analytically in terms of the full set of model parameters $\st$, $C_0$, $S$, $a$ and $H_0$ using the method in~\citet{Gus11,Gus16b}.
The calculation is outlined in Appendix~\ref{sec:smallKuExpansion}.
The resulting mean helicity becomes to second order in $\ku$:
\begin{align}
\langle H\rangle&=
H_0
+\frac{3\ku^2\st}{4(5C_0^2-27(1+2\st)(5S+3\st))(10C_0^2-27(1+\st)(10S+3\st))^2}
\bigg\{
\nn\\&
-(25+6H_0^2)C_0\abar\Big[50C_0^4+135C_0^2(10S(\st-2)+3\st(5\st-1))
\nn\\&
\hspace{2.9cm}-729(50S^2(\st-1)+18\st^3+15S\st(5\st-1))\Big]
\nn\\&
+30H_0\Big[\abar^2(90C_0^4\st-486C_0^2(5S\st-3\st^3))+50C_0^4(10S+9\st)
\nn\\&
\hspace{1.3cm}-675C_0^2(40S^2+42S\st+9\st^2)+729(5S+3\st)(10S + 3\st)^2\Big]
\bigg\}\,,
\eqnlab{meanHsmallKu}
\end{align}
where $\abar= \tilde a/\eta_0 $ (velocity and position are made dimensionless in terms of $u_0$ and $\eta_0$). \Figref{smallKu}{\bf a} shows the analytical solution for the mean helicity together with data for $\ku=10$.
In order to compensate for the different magnitudes of relevant time scales in flows with $\ku=10$ and flows with small values of $\ku$,
the parameters $\ku$ and $\st$ which depend on the correlation time of the flow have been rescaled in \Eqnref{meanHsmallKu}.
We found that multiplying the Stokes number by $4$ and the Kubo number by $1/9$ gives qualitative agreement (the effect of the former scaling is a horizontal shift of all curves and the effect of the latter scaling is a constant prefactor of the deviation from $H_0$).
\Figref{smallKu}{\bf a} shows that the small Kubo results have the same trends as those of the DNS in \Figref{Isotropic_Helicoids_Delta_H}{\bf b} and the stochastic model in \Figref{4}{\bf b}.
This shows that the trends shown in \Figref{Isotropic_Helicoids_Delta_H} are robust, they do not depend on the particular nature of the underlying flow.
Using this observation, we can use the theoretical solution of the stochastic model to get an estimate of the parameter dependence of preferential sampling of helicity for general values of the five model parameters.
Two examples of this dependence are illustrated in \Figref{smallKu}{\bf b} and {\bf c}.
\Figref{smallKu}{\bf b} shows how the observed Stokes-dependent preferential sampling of helicity depends on $C_0$.
For not too small values of $|C_0|$ the preferential sampling is similar to that observed in \Figref{Isotropic_Helicoids_Delta_H}{\bf b} and {\bf d}: co-chiral helicoids oversample helicity for small Stokes numbers, while anti-chiral helicoids oversample helicity for large Stokes numbers.
Helicoids with small $|C_0|$ behave similar to neutral particles and undersample helicity.
Similar trends are observed in a neutral flow (see \Figref{smallKu}{\bf c}).
In a neutral flow helical structures of opposite signs are equally likely, which imposes a symmetry under the simultaneous change of $H$ and $C_0$ to $-H$ and $-C_0$. This symmetry is clearly seen in \Figref{smallKu}{\bf c}: upon changing $C_0$ to $-C_0$ the average helicity changes sign.
As a consequence of this symmetry neutral particles ($C_0=0$) in neutral flows may not show preferential sampling of helicity, there is a thin line of no preferential sampling at $C_0=0$ in \Figref{smallKu}{\bf c}.

\noindent
We end the discussion on small Kubo numbers by remarking that we have applied the perturbation theory developed in Appendix~\ref{sec:smallKuExpansion} to calculate the Lyapunov exponents and Lyapunov dimension of particle clustering for small values of $\ku$, similar to the expansions for the Lyapunov exponents of spherical particles~\citep{Gus11,Gus16b}. The theory relies on the additional constraint $\st\ll\st_-$ in order for caustic singularities to be rare. In this limit we observe good agreement between theory and numerical simulations of the stochastic model (not shown).

\section{Discussion and conclusions}
\label{sec:Conclusion}
\noindent
We have presented a series of numerical and theoretical results concerning the properties of turbulent flows under strong multi-scale helical injection. We performed direct numerical simulations of the NSE
up to resolution $512^3$ and at changing the exponent of the power-law helical injection, in the limit of white-in-time noise. We first showed that there exists three different regimes for the forward energy and helicity nonlinear transfers: (i) when both transfers are directed toward small scales and the external multi-scale injection is negligible, leading to a $-5/3$ spectrum for both energy and helicity; (ii) when the energy cascade is fully nonlinear and helicity is dominated by the forcing; and (iii) when both cascades are dominated by the forcing at all scales. Finally, let us note that the theoretical prediction (\ref{eq:3spectra}) is qualitatively well reproduced by our DNS results as shown in \Figref{1}. Nevertheless, we must stress that for the dominant regime (case III in \Tabref{param}) the power law is not extremely clean. For the latter case, it would in future studies be important to extend the numerical resolution, such as to reduce spurious sub-leading terms.

For the case of turbulence under condition (III) and for a surrogate stochastic flow we studied the evolution of isotropic helicoids, presenting a systematic assessment of preferential sampling and small-scale fractal clustering for helicoids with different properties.
In particular, we showed that a suitable tuning of the chirality of the helicoids may lead to particles that behave either as being {\it lighter} or {\it heavier} than the surrounding flow.
The comparison between the turbulent and stochastic model shows very similar degree of preferential sampling for all parameters considered.
Due to the different nature of the flows, this implies that the studied preferential sampling is mainly a kinematic effect: it depends on the dynamical equations \eqnref{eqm_isotropic_helicoid} rather than on the existence and evolution of flow structures.
This suggests that the observed effects are robust to changes in the details or nature of the flow, with interesting applications also at low or moderate Reynolds numbers.
Other observables such as large-scale clustering or higher-order moments of helicity are likely to have a stronger dependency on the flow properties.

At a first glance the numerical data observed in Figs. \figref{Isotropic_Helicoids_Delta_H} and \figref{5} have sensitive and complicated parameter dependence.
Nevertheless, it is remarkable that the crude approximations \eqnref{caseone_mot_simple} and \eqnref{casetwo_simple} allow us to give a first qualitative hint about the helicoid properties in terms of the relatively simpler dynamics of spherical particles, using the {\it effective} parameters $\st_{\rm eff}$ \eqnref{St_eff}, $\beta_{\rm eff}$ \eqnref{beta_eff} and $\st/\st_+$.
The only term that is odd in $c$ or $C_0$ in these effective parameters, as well as in \Eqnref{compressibility},  is proportional to $ac C_0$.
This implies that helicoids with opposite helicoidality behave more differently the larger $ac C_0$ is.
For the flows and helicoids considered in this paper we have $ac C_0\sim 5$.
This implies an estimated helicoid size  $\tilde a$ larger than the smooth scale (approximately $10\eta$), where the point-particle approximation may not be fully correct anymore. A fully systematic analysis of finite-size effects for the helicoid properties is still lacking.
This apparent problem can be resolved by constructing helicoids with large effective $\tilde a$ while the size of the particle interacting with the fluid remains small, for example by attaching small but heavy satellite particles to the helicoid~\citep{Gus16a}.
Another solution is to consider flows or particles with larger values of $c$ or $|C_0|$ (such that $ac C_0$ is significant for small particle sizes).
On one side, the magnitude of $ac C_0$ determines how much helicoids with opposite helicoidality are different. On the other side, preferential sampling is also strongly dependent on the value of $\beta_{\rm eff}$, as exemplified by comparing $|C_0|=5$ and $|C_0|=1.6$ in \Figref{Isotropic_Helicoids_Delta_H}.
By optimizing $\beta_{\rm eff}$ it is possible to find helicoids with smaller values of $a$ that behave as light particles, i.e. have $\beta_{\rm eff}>1$.
We conclude by remarking that the construction in \Figref{IsotropicHelicoids} is just one possible way to construct isotropic helicoids and what is the most general isotropic structure which breaks mirror symmetry with a given set of parameters $C_0$, $S$, and $a$ remains an open question.\\
\noindent
The research leading to these results has
received funding from the European Union's Seventh Framework Programme
(FP7/2007-2013) under grant agreement No.  339032.
K.G. acknowledges funding from the Knut and Alice Wallenberg Foundation, Dnar. KAW 2014.0048. The numerical computations used resources provided by C3SE and SNIC.

\appendix
\section{Stochastic model for helical turbulence}
\label{sec:StochasticModel_SI}
\noindent
To construct the incompressible, homogeneous and isotropic stochastic velocity field, $\ve u=\ve\nabla\times\ve A$, used in the article, we generate the components of the vector potential $\ve A$ as a Fourier sum
\begin{align}
A_i(\ve{r},t)=\frac{(2\pi)^{3/4}}{\sqrt{3(1+\mu)}}\frac{\eta_0^{5/2}u_0}{L^{3/2}}\sum_{\sve k}\sum_{j=1}^3[(h_{j,\sve k}^-)^*a_{j,\sve{k}}(t){h_{i,\sve k}^-}+\mu(h_{j,\sve k}^+)^*a_{j,\sve{k}}(t){h_{i,\sve k}^+}]e^{{\rm i}\sve{k}\ccdot\sve{r}-\frac{k^2\eta_0^2}{4}} \,.
\eqnlab{FourierSum}
\end{align}
Here $L=10$ is the system size (we use $L=10\eta_0$ in our simulations), the wave vector $\ve k$ is summed over the components $k_j=2\pi n_j/L$ with $n_j=-20,-19,\dots,+20$ and $j=1,2,3$.
For each $\ve k$, the vector of Fourier coefficients, $\ve a_{\sve{k}}(t)$, has been expanded in terms of the eigenmodes $\ve h^\pm_{\sve k}$ of the curl operator, weighted by a factor~$\mu$ to give a bias to positive helical modes if $\mu>1$, and to negative helical modes if $0\le \mu<1$.
The coefficients $a_{i,\sve{k}}(t)$ are complex random Gaussian numbers fulfilling the condition $a^*_{i,\sve k}=a_{i,-\sve k}$ and having the statistics
\begin{align}
\langle a_{i,\sve k}(t)\rangle=0\hspace{0.5cm}\mbox{and}\hspace{0.5cm}\langle a_{i,\sve k_1}(t_1)a^*_{j,\sve k_2}(t_2)\rangle=\delta_{ij}\delta_{\sve k_1\sve k_2}e^{-|t_1-t_2|/\tau_0}\,.
\eqnlab{aCoeffsStatistics}
\end{align}
The exponential time correlation in \eqnref{aCoeffsStatistics} is generated from an underlying Ornstein-Uhlenbeck processes
\begin{align}
a_{i,\sve k}(t+\delta t)=e^{-\delta t/\tau_0}a_{i,\sve k}(t)+b_{i,\sve k}(t)\,,
\end{align}
where $\delta t$ is the time step of the simulation and $b_{i,\sve k}(t)$ are independent random Gaussian numbers that are white noise in time with statistics
\begin{align}
\langle b_{i,\sve k}(t)\rangle=0\hspace{0.5cm}\mbox{and}\hspace{0.5cm}\langle b_{i,\sve k_1}(t)b^*_{j,\sve k_2}(t)\rangle=\delta_{ij}\delta_{\sve k_1\sve k_2}(1-e^{-2\delta t/\tau_0})\,.
\end{align}
The Gaussian cutoff for large $k$ in \Eqnref{FourierSum} ensures a Gaussian spatial correlation function with correlation length $\eta_0$.
When $L\gg\eta_0$, \Eqnref{FourierSum} implies the correlation function
\begin{align}
\langle A_i(\ve r_1,t_1)A_j(\ve r_2,t_1)\rangle=\frac{\eta_0^2u_0^2}{6} e^{-|\sve r_1-\sve r_2|^2/(2\eta_0^2)-|t_1-t_2|/\tau_0}\,.
\eqnlab{Acorr}
\end{align}
From this correlation function the statistics of $\ve u$ and its spatial derivatives follows.
\noindent
To obtain the distribution $P_{0}(H)$ of helicity $H=2\ve u\ccdot\ve\Omega$ for the stochastic flow in \Eqnref{FourierSum}, we start from the joint distribution of $\ve u$ and $\ve\Omega$
\begin{align}
P = \frac{1}{8\pi^3\sqrt{\det\ma C}} e^{-\sve X{^{\rm T}}\ma C^{-1}\sve X}\,,
\end{align}
where $\ve X=(u_1,u_2,u_3,\Omega_1,\Omega_2,\Omega_3){^{\rm T}}$ and $\ma C$ is the corresponding covariance matrix (velocity is made dimensionless in terms of $u_0$ and position in terms of $\eta_0$)
\begin{align}
C_{ij}=\langle X_iX_j\rangle=
\frac{1}{12}
\left[
\begin{array}{cccccc}
4 & 0 & 0 & 2H_0 & 0 & 0\cr
0 & 4 & 0 & 0 & 2H_0 & 0\cr
0 & 0 & 4 & 0 & 0 & 2H_0\cr
2H_0 & 0 & 0 & 5 & 0 & 0\cr
0 & 2H_0 & 0 & 0 & 5 & 0\cr
0 & 0 & 2H_0 & 0 & 0 & 5
\end{array}
\right]
\end{align}
obtained from \Eqnref{Acorr} with $H_0 \equiv \sqrt{2/\pi}8(\mu^2-1)/(3(\mu^2+1))$.
After a change of coordinates $\Omega_z=(H/2-\Omega_x u_x-\Omega_y u_y)/u_z$ and integration over $\Omega_x$ and $\Omega_y$ the remaining joint distribution of $H$, $u_x$, $u_y$ and $u_z$ depends only on $H$ and the combination $\sqrt{u_x^2+u_y^2+u_z^2}$. Changing to spherical coordinates in $\ve u$-space and integrating them away gives the final distribution of helicity, \Eqnref{PH0}:
\begin{align}
P_0(H) & = \frac{9}{\pi}\frac{ |H|\exp\left[\frac{3H_0H }{5-H_0^2}\right]K_1\left[\frac{3\sqrt{5}|H|}{5-H_0^2}\right] }{\sqrt{5\left[5-H_0^2\right]}}\,,
\end{align}
where $K_\nu(x)$ is the modified Bessel function of the second kind. The average helicity of the flow is determined from the helicity bias $\mu$ as follows:
\begin{align}
\langle H\rangle_{\rm flow}=\int_{-\infty}^\infty{\rm d}H H P_0(H)=H_0=\frac{8}{3}\sqrt{\frac{2}{\pi}}\frac{\mu^2-1}{\mu^2+1}\,.
\end{align}

\section{Expansion around deterministic trajectories}
\label{sec:smallKuExpansion}
\noindent
Here we outline the series expansion used to calculate the mean helicity \eqnref{meanHsmallKu} of isotropic helicoids for small values of $\ku$ in the statistical model.
The expansion follows the method introduced in~\citet{Gus11} and reviewed in~\citet{Gus16b}.
We want to expand the dynamics in the dimensionless equations of motion (\Eqnref{eqm_isotropic_helicoid_matrix} together with $\dot{\ve r}=\ku\ve v$) around the deterministic solution $\ve r^{(\rm d)}$ obtained without flow, i.e. when $\ve u=\ve\Omega=\ve 0$.
Since we in this work only consider homogeneous steady-state statistics, we can put all initial conditions to zero for simplicity.
We therefore expand around the simple deterministic solution $\ve r^{(\rm d)}=\ve 0$.
\noindent
An implicit solution to the dynamics in \Eqnref{eqm_isotropic_helicoid_matrix} together with $\dot{\ve r}=\ku\ve v$ can be found by first solving the diagonal equations for $\ve\zeta$ in Eqs.~\eqnref{diag1} and \eqnref{diag2}, then transforming back to $\ve v$ and finally integrating to obtain $\ve r$.
We find that the following is an exact implicit solution to the dynamics:
\begin{align}
\ve r_t&=-\ku\int_0^t{\rm d}t_1\int_0^{t_1}{\rm d}t_2\left[e^{\st_+(t_2-t_1)/\st}\ve U_+(\ve r_{t_2},t_2)+e^{\st_-(t_2-t_1)/\st}U_-(\ve r_{t_2},t_2)\right]\,
\eqnlab{Expansion_Trajectory}
\end{align}
where
\begin{align}
\ve U_\pm=\frac{\st_\pm}{\st_\pm-\st_\mp}\frac{1}{\st}\left[(\st_\mp-1)\ve u-\frac{2C_0a}{9}\ve\Omega\right]\,.
\end{align}
A series expansion of the flow velocity (and spatial derivatives thereof) around the deterministic trajectory $\ve r^{(\rm d)}=0$ gives
\begin{align}
u_i(\ve r_t,t)=u_i(\ve 0,t)+\frac{\partial u_i}{\partial r_j}(\ve 0,t)r_{j,t}+\frac{1}{2}\frac{\partial^2 u_i}{\partial r_j\partial r_k}(\ve 0,t)r_{j,t}r_{k,t}+\dots\,.
\eqnlab{Expansion_Flow}
\end{align}
\Eqnref{Expansion_Flow} is an expansion of the flow velocity in terms of the displacement from $\ve r=0$.
Recursively substituting $\ve u(\ve r_t,t)$ (and derivatives thereof) from \Eqnref{Expansion_Flow}, and $\ve r_t$ from \Eqnref{Expansion_Trajectory} into  \Eqnref{Expansion_Flow}, we obtain an increasingly refined approximation of the flow evaluated along the true trajectory $\ve r_t$.
Since $\ve r_t$ is of order $\ku$ in \Eqnref{Expansion_Trajectory}, we can use $\ku$ to keep track of the order of $\ve r_t$ in the expansion.
Truncating the recursive expansion of \Eqnref{Expansion_Flow} at some order in $\ku$, one obtains the approximate expression for $\ve u$ along a trajectory $\ve r_t$ to this order in $\ku$.
To evaluate an approximation for the helicity along a particle trajectory, we do a similar expansion for $\ve\Omega(\ve r_t,t)$ to form $H=2\ve u(\ve r_t,t)\ccdot\ve\Omega(\ve r_t,t)$.
Finally, we evaluate the steady-state average $\langle H\rangle=2\langle\ve u(\ve r_t,t)\ccdot\ve\Omega(\ve r_t,t)\rangle$, where the average is taken over an ensemble of trajectories and can be explicitly evaluated for the stochastic model in terms of the known Eulerian correlation function in \Eqnref{Acorr}, see~\citet{Gus11,Gus16b} for more details.
As a result, we obtain the expression in \Eqnref{meanHsmallKu}.

\bibliographystyle{jfm}

\end{document}